\documentclass[fleqn,usenatbib]{mnras}

\usepackage[T1]{fontenc}
\usepackage{ae,aecompl}

\usepackage{graphicx}	
\usepackage{amsmath}	
\DeclareMathOperator{\sech}{sech}

\usepackage{amssymb}	
\usepackage{pdflscape}

\usepackage{txfonts} 

\usepackage{epstopdf}
\usepackage{journals}

\usepackage[none]{hyphenat}
\graphicspath{{images_aj_formatted/}}
\DeclareGraphicsExtensions{.pdf,.png,.jpg, .jpeg}
\usepackage{natbib}
\usepackage{xcolor} 
\usepackage[normalem]{ulem}
\usepackage{xspace} 

\usepackage[english]{babel} 
\usepackage{threeparttable}

\newcommand{\rpm}{\sbox0{$1$}\sbox2{$\scriptstyle\pm$}
	\raise\dimexpr(\ht0-\ht2)/2\relax\box2 }

\title[X-shaped bulge photometric model]{New X-shaped bulge photometric model as a tool for measuring B/PS bulges and their X-structures in photometric studies}
\author[Anton A. Smirnov, Sergey S. Savchenko]
{Anton A. Smirnov$^{1,2}$\thanks{E-mail:
zeleniikot@gmail.com}, Sergey S. Savchenko$^{1,3}$\\
$^{1}$St. Petersburg State University,
Universitetskij pr.~28, 198504 St. Petersburg, Stary Peterhof, Russia\\
$^{2}$Central (Pulkovo) Astronomical Observatory of RAS, Pulkovskoye Chaussee 65/1, 196140 St. Petersburg, Russia\\
$^{3}$ Special Astrophysical Observatory, Nizhniy Arkhyz, 369167 Karachai-Cherkessia, Russia\\
}

\date{Accepted XXX. Received YYY; in original form ZZZ}

\pubyear{2020}

\begin{document}
\label{firstpage}
\pagerange{\pageref{firstpage}--\pageref{lastpage}}
\maketitle

\begin{abstract}
  Recent orbital studies of 3D bar structure in various numerical and analytical models show that X-structures that
  reside in boxy/peanut-shaped (B/PS) bulges are not delineated by some specific type of orbits, but are natural parts
  of them and formed by the same orbits that constitute such bulges. This implies that to accurately account for B/PS
  bulges and their X-structures in photometric studies, one needs the photometric model of B/PS bulge that includes an
  X-structure as its natural part. To find such a model, we considered a self-consistent numerical galaxy model where a
  typical B/PS bulge arises. Using spectral characteristics of particle-``stars'', we decomposed the galaxy model onto
  the bar and non-bar components. We used the extracted 3D bar component to find an appropriate B/PS bulge
  photometric model, which can account for X-structures residing in such bulges. The resulted B/PS bulge photometric
  model has a truncated 2D Sersic profile with truncations introduced above (in the upper half-plane) and below (in the
  bottom half-plane) the rays of X-structures. We applied this model to represent B/PS bulges of various numerical
  models and some real galaxies. The comparison with previous works revealed that there are systematic shifts between the
  X-structure parameters of the same galaxies measured within the different approaches. We found that the geometric
  parameters of X-structures of real and modelled galaxies are consistent with each other if we measure them using our
  new model.
\end{abstract}

\begin{keywords}
galaxies: bulges -- galaxies: bar  -- galaxies: photometry -- galaxies: structure.

\end{keywords}

\section{Introduction}
X-structures are cross-shaped structures observed in some disk galaxies seen edge-on. The classic example of the galaxy
with such a peculiar morphology is NGC~128. The central area of this galaxy has a peanut-like profile. A careful
observer can identify four thick lobes protruding from the disc plane and distinguishable on the background of the
peanut. These lobes are almost symmetric with respect to the disc plane, and they are connected to the very central area
of the galaxy. Such structures frequently referred as an X-shape or an X-structure. The whole peanut-like structure
(including X-structure) is usually referred as a peanut bulge. In some galaxies (for example, NGC~7332) the bulges of a
similar physical nature appear in a form of the box. The common wisdom is to refer to the whole set of such objects as
boxy/peanut-shaped (B/PS) bulges. 
\par 
Studies of B/PS bulges in disc galaxies have a rather long history. First seminal work where such bulges were
distinguished was done in the late fifties by~\cite{Burbidge_Burbidge1959}. Authors considered three galaxies one of
which was mentioned above NGC~128. In this work one can also find the first implicit reference to the X-structure
inhabiting this galaxy: ``\textit{At the widest parts of the nuclear region there are four bulges of about equal size,
  coming out of the nucleus itself like a cross.}''~\citep{Burbidge_Burbidge1959}. The term `X-shape' itself was
introduced later on in the study of IC~4767 galaxy by~\cite{Whitmore_Bell1988}. The isophotes in the central part of
this galaxy show a typical peanut profile . The `X-shape' itself was revealed after subtracting the disc model
from the original image of the galaxy.
\par 
More recent works~\citep{Jarvis1986,Shaw1987,deSouza_DosAnjos1987, Lutticke_etal2000,Erwin_Debattista2013,
  Yoshino_Yamauchi2015, Erwin_Debattista2017,Kruk_etal2019} show the fraction of disc galaxies with B/PS bulges is quite
significant. The fraction varies from 20\%-25\%~\citep{Shaw1987,deSouza_DosAnjos1987,Yoshino_Yamauchi2015} to
40\%~\citep{Lutticke_etal2000} for different samples. \cite{Kruk_etal2019} tried to account for the projection effects
and obtained that 70\% of disc galaxies should posses B/PS bulges to the epoch $z=0$ (present day). These studies
generally show that B/PS bulges are quite common inhabitants of disc galaxies.
\par 
Historically, there were different hypothesises concerning the physical nature of B/PS bulges and X-structures. Among
them are the collapse of the protogalactic cloud~\citep{Jarvis1981}, the galaxy merging~\citep{Hernquist_Quinn1988} and
the accretion of the material during tidal encounters~\citep{Schweizer_Seitzer1988}.  Although not
completely~\citep{Binney_Petrou1985}, these theories mainly came to halt with the work of \citet{Combes_Sanders1981}
where the authors showed that the bars thickened in the vertical direction take the form of a peanut or a box depending
on the angle between the bar major axis and the observer line of sight (LoS), ${\approx} 90^\circ$ for a peanut to be
observed and generally less than 50$^\circ$ for a box to be observed, respectively\footnote{It is difficult to determine
  a strict value of the angle in the general case since it depends on the size of the peanut itself.}. Further numerical
studies reinforced these results~\citep{Combes_Sanders1981, Combes_etal1990, Friedli_Pfenniger1990, Raha_etal1991,
  Pfenniger_Friedli1991}. Subsequent kinematic studies of stars and ionised gas showed that B/PS bulges rotate
cylindrically~\citep{Bertola_Capaccioli1977,Kormendy_Illingworth1982} and have a specific shape of LoS velocity
distribution that arising due to bars residing in the observed galaxies~\citep{Kuijken_Merrifield1995,
  Bureau_Freeman1999, Merrifield_Kuijken1999, Veilleux_etal1999, Chung_Bureau2004}. These facts also supported the idea
that B/PS bulges are just thickened bars. \textcolor{black}{Recent studies have established a more accurate relationship between the B/PS bulges and their bars. \cite{Laurikainen_etal2014, Athanassoula2015} showed that the so-called ``barlenses'' (near-circular central parts of the bars found in some face-on galaxies) seem to be the manifestations of the B/PS bulges when the latter viewed from near face-on orientations. \cite{Laurikainen_Salo2017} provided evidence for the existance of such a link by a systematic study of how B/PS bulges and barlenses metric properties vary as a function of viewing inclination both in real and modelled galaxies.}

\par
The detailed reasons for bars to thicken are still the subject of debates. There are three possible physical mechanisms
including instability of hose-pipe type~\citep{Raha_etal1991}, the resonance trapping of individual
orbits~\citep{Quillen2002}, and so-called resonance heating~\citep{Quillen_etal2014}. The latter two are similar in a
sense that the basic agent that forces stars to jump out from the disc plane is 2:1 vertical inner Lindblad resonance
(vILR). Stars at the resonance experience non-zero net effect from the oscillating bar potential. They gradually
accumulate small changes in momentum and sooner or later are forced to jump out of the disc plane. The actual
distinction between trapping and heating is that the heating model is trying to account for the bar slow
downing~\citep{Athanassoula2005} that leads to a change of the vILR location. The hose-pipe instability is associated
with the bending waves. They become unstable if there is no sufficient ratio of vertical to radial velocity dispersion,
$\sigma_z/\sigma_R$, to prevent the waves growth~\citep{Toomre1966, Poliachenko_Shukhman1977, Araki1985,
  Merritt_Sellwood1994}.  An interesting result concerning the mechanisms was obtained by
\cite{Pfenniger_Friedli1991}. The authors run the experiments with forced vertical symmetry and showed that the peanut
still forms, but on longer time scales. This result supports the idea that both the resonance trapping and bending waves
are important for the growth of the peanut (see also recent work by~\citealt{Sellwood_Gerhard2020}).
\par
The physical nature of the X-structures themselves is also an open question, although significant progress has already
been made in this direction. The spatial resolution of the models was not very good in the first numerical studies of 3D
bar structure (${\approx} 10^3-10^4$ particles represented the disc component). X-structures were only distinguished in
unsharp masked images constructed from such models. Some authors~\citep{Friedli_Pfenniger1990,Pfenniger_Friedli1991}
therefore suggested that X-structures are akin to an optical illusion due to the tendency of eyes to perceive the intensity
gradients instead of actual intensity values. However, with an increase of spatial resolution, it became evident that the
X-structures are real density enhancements that can be observed even without unsharp masking processing
(see~\cite{Smirnov_Sotnikova2018} for many representative examples). The question then is why such density enhancements
are observed. Studies of orbit composition of B/PS bulges and X-structures in different numerical and analytical
models~\citep{Patsis_etal2002,Quillen2002,Patsis_Katsanikas2014a,Quillen_etal2014,Parul_etal2020} showed that an X-shape
is observed in these models due to a tendency of a star to spend more time near turning points of its trajectory. More
specifically, 3D bars are constituted by different types of periodic, quasi-periodic, and sticky chaotic
orbits~\citep{Pfenniger_1984,Pfenniger_Friedli1991, Skokos_etal2002, Patsis_etal2002, Patsis_Katsanikas2014a,
  Patsis_Katsanikas2014b, Patsis_Harsoula2018, Patsis_Athanassoula2019}. Stars that move along such orbits spend
different periods of time in different parts of their trajectories. For example, stars moving along banana-shaped
orbits~\citep{Pfenniger_Friedli1991} spend more time at the highest points of their trajectory~\citep{Patsis_etal2002,
  Patsis_Katsanikas2014a}. Therefore, the bulk of such orbits produces a density profile with visible density
enhancements at the highest points of such orbits. For an X-structure to be observed, these density enhancements should
be aligned along an almost straight line for orbits with different apocentric distances. This is indeed the case for the
realistic bar potential~\citep{Patsis_etal2002, Quillen2002, Patsis_Katsanikas2014a,
  Quillen_etal2014}. \cite{Parul_etal2020} showed that the orbits of a more complicated morphology than banana-shaped
orbits can build an X-structure in a similar manner. In general, cited works showed that X-structures and B/PS bulges
are produced by the same orbits. They are not constituted by different types of orbits like, for example, disc and
classical bulge components. The open question that has to be answered in the upcoming studies is what types of orbits
are actually presented in real galaxies.
\par 
An important aspect of the physical nature of B/PS bulges (and their X-structures) is that they are products of a
secular evolution of the host galaxy. Consequently, the physical properties of these components should be tightly
connected with the properties of the underlying gravitational potential. This includes the potential imposed by the dark
matter component~\citep{Smirnov_Sotnikova2018}. Therefore, the detailed observational studies of B/PS bulges and
X-structures can provide additional important constraints on the dynamics of the galaxies that host such objects.
\par
The most promising candidates for such studies are galaxies seen edge-on. In such galaxies, one can try to assess not only
the sizes of the B/PS bulges (which can be done in intermediate inclined galaxies, see~\citealt{Erwin_Debattista2013})
but characterise the shape of the B/PS bulges~\citep{Ciambur_Graham2016} and characterise their most prominent features
- X-structures~\citep{Laurikainen_Salo2017,Savchenko_etal2017}. An important thing to stress out here is that all groups
of authors used different processing algorithms. \cite{Ciambur_Graham2016} quantified B/PS bulges as a whole (including
X-structures) fitting isophotes by a Fourier modified ellipses, while \cite{Laurikainen_Salo2017, Savchenko_etal2017}
studied only the X-structures using unsharp-masking and photometric decomposition, respectively. The difference in
approaches gives rise to some inconsistency in results obtained in these works. The most clear example is ESO~443-042
galaxy. \cite{Ciambur_Graham2016} obtained that an X-structure residing in this galaxy is very flattened (their horizontal and
vertical sizes \textcolor{black}{are 20.0 arcsec and 5.1 arcsec}, respectively, see their table~2), while the X-structure extracted
by~\cite{Laurikainen_Salo2017} seems quite typical (horizontal and vertical sizes 15.7 arcsec and 12.0 arcsec,
respectively, see their table~F.2).
\par 
In our previous work where we considered a rather large set of different numerical models~\cite{Smirnov_Sotnikova2018} we
obtained that modelled X-structures seem to be less flattened than real observed X-structures. We characterised the
flatness of X-structure in terms of the value of the angle between the disc plane and the ray of an
X-structure. X-structures in our models demonstrated opening angles between 26$^\circ$ and 42$^\circ$, while the opening
angles of real X-structures have values in nearly the same range with a slight shift to the lower
values~\citep{Savchenko_etal2017}. This is disturbing because the opening angles observed in real galaxies actually
should be shifted to the greater values than 30$^\circ$ due to the projection effects (see figure~22 in
\cite{Smirnov_Sotnikova2018}). In principle, such inconsistency can also arise due to slightly different approaches to
measuring X-structure opening angles in \cite{Smirnov_Sotnikova2018} and
\cite{Savchenko_etal2017}. In~\cite{Savchenko_etal2017} the density maxima were measured along photometric cuts
perpendicular to the rays of X-structure while in \cite{Smirnov_Sotnikova2018} we used cuts parallel to the disc
plane. However, this should be strictly verified because there are exist some physical reasons why the observed
X-structures can appear more flattened. For example, X-structures flatten in course of their secular
evolution~\citep{Smirnov_Sotnikova2018}. Therefore, a large flatness of observed X-structures can be an indicator that
they live on average longer than 6-8 Gyrs~\citep{Smirnov_Sotnikova2018}. Strongly flattened X-structures can be also
observed in galaxies with a large contribution of dark matter within the optical radius of the
disc~\cite{Smirnov_Sotnikova2018}.
\par
One more inconsistency between~\cite{Smirnov_Sotnikova2018} and \cite{Savchenko_etal2017} is that X-structures extracted
by~\cite{Savchenko_etal2017} usually show curved rays (see figure~4 there) which is not the case for X-structures from
numerical models. Again, this can be can an indicator that there are a mix of orbital families which support B/PS bulge in
observed galaxies \citep{Patsis_Harsoula2018,Parul_etal2020} or the effect due to the difference in processing
algorithms.
\par 
In the present work, we would like to specifically address the question of whether or not the mentioned inconsistencies are
due to the difference in processing algorithms or a real physical effect. To this aim we apply exactly the same methods
of B/PS bulges analysis for some numerical models from~\cite{Smirnov_Sotnikova2018} and real galaxies from the samples
considered by \cite{Ciambur_Graham2016} and \cite{Savchenko_etal2017}. \textcolor{black}{We note that the idea to study the B/PS bulges of numerical models and real galaxies using the same processing algorithms has already been exploited by~\cite{Laurikainen_Salo2017} and allowed the authors to obtain important results concerning B/PS bulges -- barlenses connection, thus proving to be quite fruitful.}
\par 
We choose to work within the framework of photometric decomposition as it was done in~\citep{Savchenko_etal2017}. But
unlike the photometric model for the X-structure considered in~\cite{Savchenko_etal2017}, we tried to construct a new
photometric model which accounts simultaneously for both B/PS bulge and its X-structure. Such an approach is more
consistent with recent studies of X-structures and B/PS bulges where authors showed that the X-structures are
inextricable parts of the B/PS bulges~\citep{Patsis_etal2002,Patsis_Katsanikas2014a,Parul_etal2020}. We deduced the new
photometric model on the basis of ``dynamical'' decomposition of a numerical model of a Milky Way-like galaxy from
\cite{Smirnov_Sotnikova2018}. The ``dynamical'' decomposition is the decomposition of the galaxy onto different orbital
groups or families based on the orbital frequencies. This decomposition was done using the methods of spectral
dynamics~\citep{Binney_Sprigel1982} as it was done in~\cite{Parul_etal2020}.
\par
The manuscript has the following structure. In Section~\ref{sec:phot_model} we present our new photometric model. In
Section~{\ref{sec:fit}} we give the details of the fitting procedure. In Section~\ref{sec:numerical_models} we study how
the new B/PS bulge photometric model changes its parameters depending on other photometric components (disc, classical
bulge), as well as projection effects (disc inclination angle, bar \textcolor{black}{azimuthal viewing angle}), and also consider some important trends arising
due to the difference in the physical parameters of the modelled galaxies. In Section~\ref{sec:real_galaxies} we
describe our sample of real galaxies and show a few examples how the new photometric model works in case of real
galaxies. In Section~\ref{sec:comp_previous_works} we compare our results for full galaxy sample with results of
previous works, and verify whether different measuring techniques give the same results on the same galaxies. In
Section~\ref{sec:comp_models} we compare B/PS bulge geometric parameters of real galaxies with those of some numerical
models. In Section~\ref{sec:sum} we summarise our results.

\section{Photometric model of the X-shaped bulge}%
\label{sec:phot_model}%


\begin{figure*}%
\begin{minipage}[t]{0.95 \textwidth}
\includegraphics[width=1.0\textwidth]{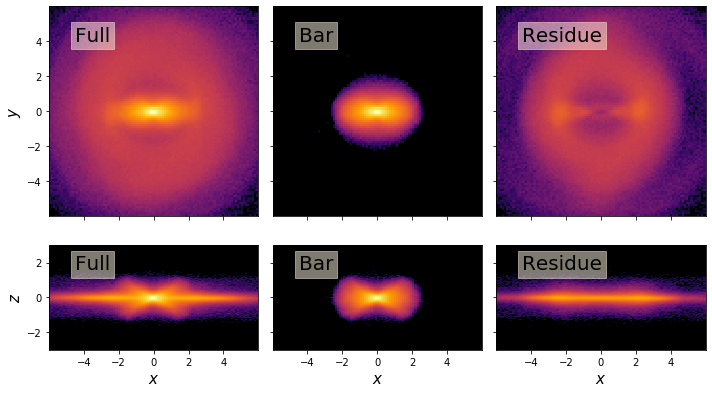}%
\end{minipage}%
\caption{The decomposition of the galaxy model (\textit{left column}) on the bar part (\textit{mid column}) and the residue part (\textit{right column}) by means of the frequency analysis.}
\label{fig:bar}
\end{figure*}

\begin{figure*}%
\begin{minipage}[t]{0.3 \textwidth}
\includegraphics[scale=0.33]{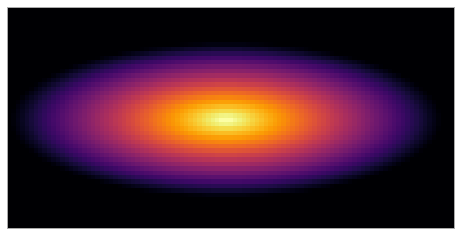}%
\end{minipage}%
\begin{minipage}[t]{0.3 \textwidth}
\includegraphics[scale=0.33]{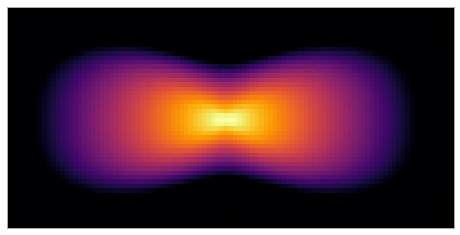}%
\end{minipage}%
\begin{minipage}[t]{0.3 \textwidth}
\includegraphics[scale=0.33]{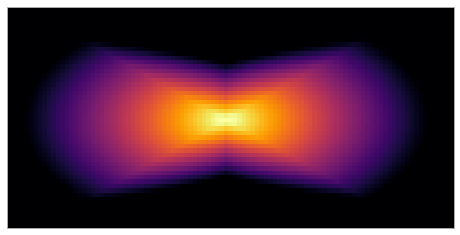}%
\end{minipage}%
\begin{minipage}[t]{0.08 \textwidth}%
\vspace{-2.8cm}
\includegraphics[scale=0.35]{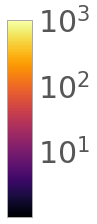}%
\end{minipage}\\%
\begin{minipage}[t]{0.3 \textwidth}
\includegraphics[scale=0.33]{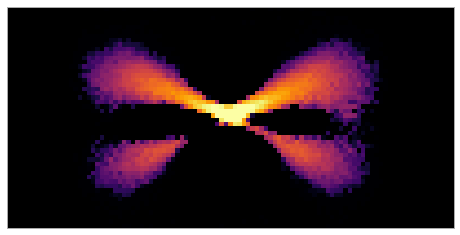}%
\end{minipage}%
\begin{minipage}[t]{0.3 \textwidth}
\includegraphics[scale=0.33]{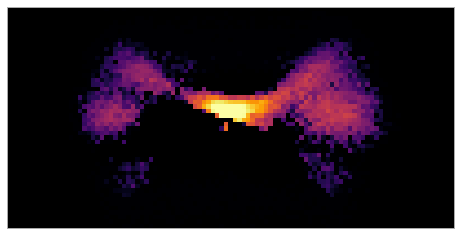}%
\end{minipage}%
\begin{minipage}[t]{0.3 \textwidth}
\includegraphics[scale=0.33]{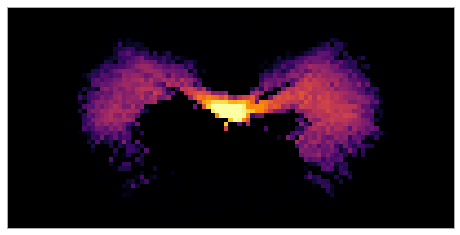}%
\end{minipage}%
\begin{minipage}[t]{0.08 \textwidth}
\vspace{-2.8cm}
\includegraphics[scale=0.35]{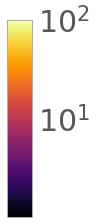}%
\end{minipage}\\%
\begin{minipage}[t]{0.3 \textwidth}
\includegraphics[scale=0.33]{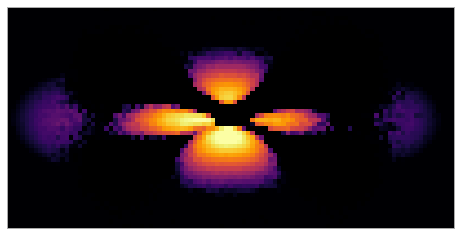}%
\end{minipage}%
\begin{minipage}[t]{0.3 \textwidth}
\includegraphics[scale=0.33]{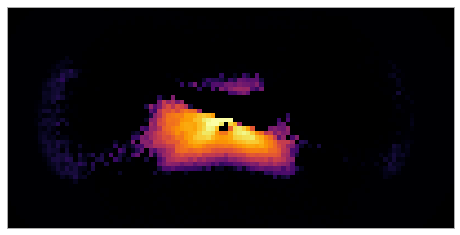}%
\end{minipage}%
\begin{minipage}[t]{0.3 \textwidth}
\includegraphics[scale=0.33]{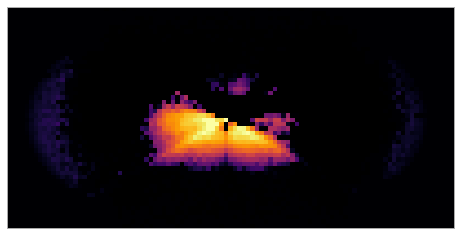}%
\end{minipage}%
\begin{minipage}[t]{0.08 \textwidth}
\vspace{-2.8cm}
\includegraphics[scale=0.35]{comp_funcs_AS_res_colorbar}%
\end{minipage}%
\caption{Different photometric models of the vertical bar distribution shown in Fig.~\ref{fig:bar} (\textit{upper row}) and the reside images, the positive part (\textit{middle row}) and the negative part (\textit{bottom row}). \textit{Left column}: Sersic profile; \textit{Middle column}: model from~\protect\cite{Savchenko_etal2017}; \textit{Right column}: the photometric model from the present work.}
\label{fig:comp_funcs}
\end{figure*}


\begin{figure}%
\begin{minipage}[t]{0.45 \textwidth}
\includegraphics[width=1.0\textwidth]{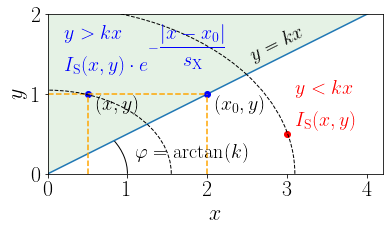}%
\end{minipage}%
\caption{Schematic image showing how the intensity is calculated in the X-shaped bulge photometric model defined by Eq.~(\ref{eq:model}). $I_S$ stands for the intensity calculated for a non-modified Sersic function.}
\label{fig:phot_model}
\end{figure}

\begin{figure*}%
\begin{minipage}[t]{0.3 \textwidth}
\includegraphics[scale=0.33]{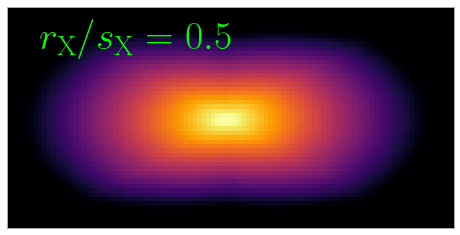}%
\end{minipage}%
\begin{minipage}[t]{0.3 \textwidth}
\includegraphics[scale=0.33]{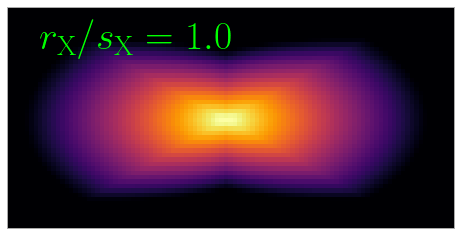}%
\end{minipage}%
\begin{minipage}[t]{0.3 \textwidth}
\includegraphics[scale=0.33]{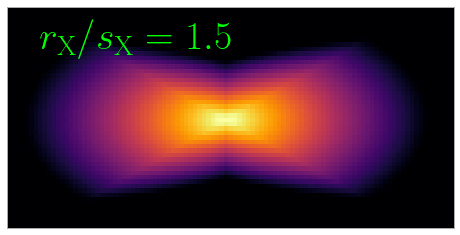}%
\end{minipage}\\%
\begin{minipage}[t]{0.3 \textwidth}
\includegraphics[scale=0.33]{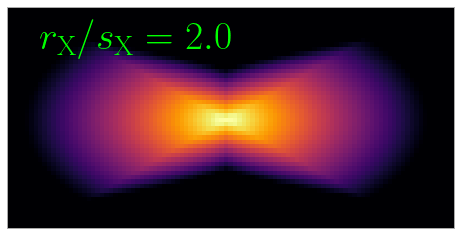}%
\end{minipage}%
\begin{minipage}[t]{0.3 \textwidth}
\includegraphics[scale=0.33]{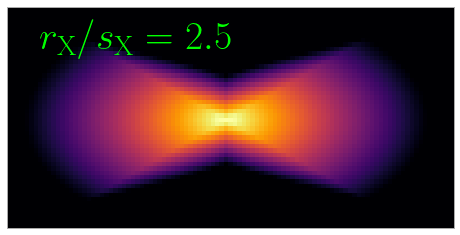}%
\end{minipage}%
\begin{minipage}[t]{0.3 \textwidth}
\includegraphics[scale=0.33]{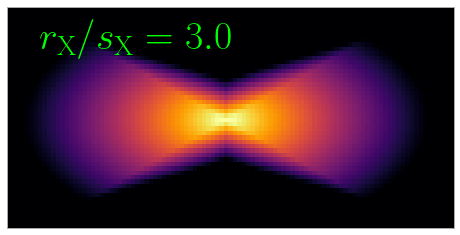}%
\end{minipage}%
\caption{Examples of the X-shaped bulge model appearance for different values of the sharpness parameter $r_\mathrm{X}/s_\mathrm{X}$ and the fixed effective length $r_\mathrm{X}$, \textcolor{black}{the Sersic index $n_\mathrm{X}=1$, and the opening angle $\varphi=30^\circ$}. The other parameters are the same for all cases .}
\label{fig:comp_sharp}
\end{figure*}

\begin{figure*}%
	\begin{minipage}[t]{0.3 \textwidth}
		\includegraphics[scale=0.33]{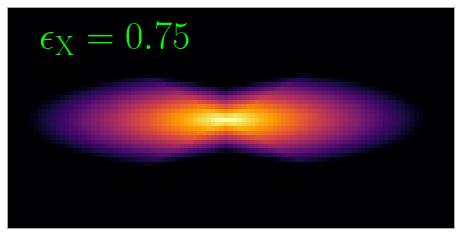}%
	\end{minipage}%
	\begin{minipage}[t]{0.3 \textwidth}
		\includegraphics[scale=0.33]{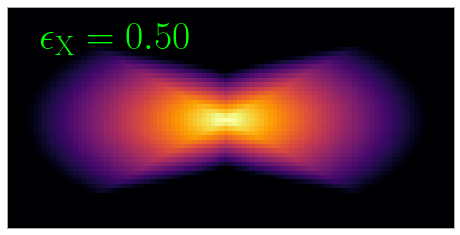}%
	\end{minipage}%
	\begin{minipage}[t]{0.3 \textwidth}
		\includegraphics[scale=0.33]{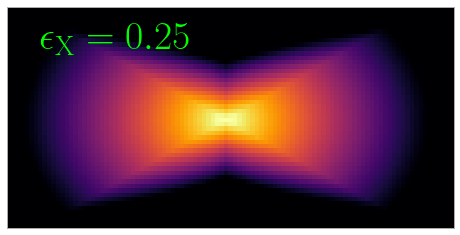}%
	\end{minipage}\\%
	\begin{minipage}[t]{0.3 \textwidth}
		\includegraphics[scale=0.33]{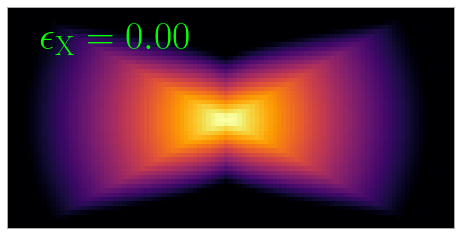}%
	\end{minipage}%
	\begin{minipage}[t]{0.3 \textwidth}
		\includegraphics[scale=0.33]{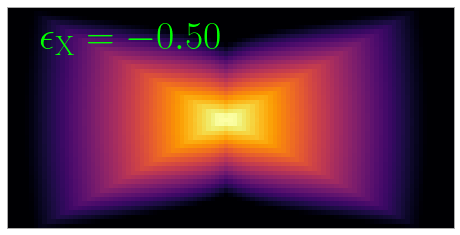}%
	\end{minipage}%
	\begin{minipage}[t]{0.3 \textwidth}
		\includegraphics[scale=0.33]{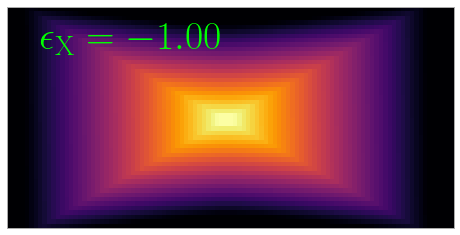}%
	\end{minipage}%
	\caption{Examples of the X-shaped bulge model appearance for different values of the ellipticity $\epsilon_\mathrm{X}$ and the fixed values of the sharpness parameter $r_\mathrm{X}/s_\mathrm{X}=2$, \textcolor{black}{the Sersic index $n_\mathrm{X}=1$, and the opening angle $\varphi=30^\circ$}. The other parameters are the same for all cases.}
	\label{fig:comp_ell}
\end{figure*}

B/PS bulges are usually described by a Sersic function~\citep{Sersic},
\begin{equation}
I(x,y)=I_e \exp \left[-\nu_n\left(\left(\frac{r}{r_e}\right)^{1/n}-1\right)\right],
\label{eq:ser}
\end{equation}
where $I_e$ is the effective brightness , $r_e$ -- the effective radius,
$n$ -- the Sersic index, $\nu_n$ is a function depending on $n$~\citep{Caon_etal1993} and $r$ corresponds to major axis of the ellipse with ellipticity $\epsilon$ that passes through the point with coordinates $(x,y)$, 
or Ferrers function~\citep{Ferrers1877},
\begin{equation}
I(r)=I_0 \left[ 1 - \left( \frac{r}{r_\mathrm{out}}\right)^{2-\beta}\right]^\gamma, r<r_\mathrm{out}
\label{eq:Fer}
\end{equation}
where $I_0$ is the central intensity, $r_\mathrm{out}$ is the truncation radius, $\beta$ and $\gamma$ control the
overall shape of the profile, and \textcolor{black}{$r$ is the same isohotal radius as in Eq.~(\ref{eq:ser}).} Both functions
lack the ability to account for the X-shaped structures in real galaxies because of their symmetries. This is a crucial
problem if we want to measure the X-structure parameters in the framework of the photometric
decomposition. \cite{Savchenko_etal2017} solved this problem by adding the Fourier distortion to the Ferrers
profile~(\ref{eq:Fer}) in the following form:
\begin{equation}
r=r_0 \left[1 + a_4 \cos(4 (\varphi + \varphi_0))\right],
\label{eq:r_dist}
\end{equation}
where $a_4$ and $\varphi_0$ are the amplitude and phase of Fourier distortion, respectively. Right part of
Eq.~(\ref{eq:r_dist}) is meant to be inserted in Eq.~(\ref{eq:Fer}) instead of $r$ variable thus obtaining the
profile with four symmetric lobes. The photometric model obtained in this way can indeed describe the B/PS bulge and its
X-structures, but rather not convenient for measuring the parameters of X-structures. For example, to obtain the opening
angle value, one needs to perform an additional round of measurements using the photometric cuts as it was done
in~\cite{Savchenko_etal2017}.
There is also a problem with photometric cuts that the result can depend on whether one uses
cuts parallel to the disc plane or at some specific inclination to the plane. Here we want to avoid this problem and
characterise B/PS bulges and their X-structures only employing an appropriate photometric model. To this aim, we
construct a new photometric model in which the value of the opening angle is set explicitly.
\par
Let us consider the bar and the corresponding B/PS bulge in some typical galaxy model where the bar naturally arises due
to effects of the self-gravity (precise details of the simulation will be given in the next section). Fig.~\ref{fig:bar}
shows the density distributions in both $(xy)$ and $(xz)$ planes produced by all particles representing the stellar
component (\textit{left column}), bar particles (\textit{middle column}) and all particles that are not the bar
particles (\textit{right column}) in the model after 6 Gyr of evolution. The presented division on the bar and not bar
components was done using the spectral characteristics of the orbits, $f_\mathrm{R}/f_x$ and $f_z/f_x$ frequency rations
and the values of $f_x$ and $f_y$ frequencies themselves, where $x,y,z$ are coordinates in the reference frame
co-rotating with the bar, $R$ --- the cylindrical radius and $f_x,f_y, f_z$, and $f_\mathrm{R}$ are the frequencies of
oscillations along the corresponding axes. The bar particles was identified as the particles that have
$f_\mathrm{R}/f_x < 4$, $f_z/f_x< 2.1$ and $f_x\gtrsim21$ km s$^{-1}$ kpc$^{-1}$ and $f_y\gtrsim21$ km s$^{-1}$
kpc$^{-1}$. Analogous division was done in~\cite{Parul_etal2020} and a similar approach was also used to distinguish the
bar by several authors before that~\citep{,Ceverino_Klypin2007,Portail_etal2015b, Gajda_2016}. Here we stray from the
``typical'' frequency ratio condition $f_\mathrm{R}/f_x=2 \pm 0.1$ used in the previous works, because we found it to be
insufficient to accurately extract the bar in this model. Since the bar shown in Fig.~\ref{fig:bar} is cleared from the
disc contribution (and also ansae, see the right column of Fig.~\ref{fig:bar}) this makes it an ideal target for
studying how well various photometric models represent the actual bar density profile. This is an advantage of the
galaxy models laying in the fact that we can avoid the need to account for other components and study only the density
distribution of the B/PS bulge itself.
\par
Fig.~\ref{fig:comp_funcs} shows three different best-fit models of the bar density distribution from Fig.~\ref{fig:bar}: pure Sersic model (\textit{left column}), model from~\cite{Savchenko_etal2017} (\textit{middle column}) and our new photometric model (\textit{right column}), details of which will be given below. The fitting procedure was done using {\small IMFIT} package~\citep{imfit}. To find the optimal values of parameters, we used Gaussian-based maximum likelihood statistic which is the total $\chi^2$ for the model:
\begin{equation}
-2 \ln \mathcal{L} = \chi^2 = \sum_{i=0}^N w_i \, (I_{m,i}-I_{d,i})^2,
\label{eq:stat}
\end{equation}
where $w_i$ are the pixel weights and $I_{m,i}$ and $I_{d,i}$ are intensities of the individual pixels in the
photometric model and in the image which we try to fit, respectively. \textcolor{black}{Hereandafter we give the chi-square values normilised to the degrees-of-freedom $\nu=n-m$, where $n$ is the number of unmasked pixels and $m$ is the number of free parameters.} The Sersic model is given in
Fig.~\ref{fig:comp_funcs} mostly for methodological purposes. It just shows that the usual Sersic model applied to the
bar gives four bright rays in the residue images, which we then call X-structures. The model of
\cite{Savchenko_etal2017} is good at capturing the overall behaviour of the intensity profile but does not have the
sharp rays, the parameters of which we can associate with the parameters of X-structures. To construct a more practical
photometric model in which the value of the opening angle would be an explicit parameter of the model, we considered a
Sersic profile~(\ref{eq:ser}) and add the truncation to it in the following form:
\begin{equation}
I(x,y) = 
\begin{cases}
I_\mathrm{S}(x,y), \, y \leq kx, \\
I_\mathrm{S}(x,y) \cdot \exp (-|x-x_0|/s_\mathrm{X}), \, \, y>kx.
\end{cases}
\label{eq:model}   
\end{equation}
Here $I_\mathrm{S}(x, y)$ stands for the Sersic intensity profile defined in Eq.~(\ref{eq:ser}), $x$ and $y$ are the pixel
coordinates, $k$ is an angular coefficient of the X-structure ray which translates into the opening angle value
$\varphi=\arctan k$, $x_0$ is the abscissa of the point that has an ordinate $y$ but belongs to the X-structure ray,
$x_0=y/k$, and $s_\mathrm{X}$ is the parameter that determines how fast the intensity is declining above the
ray. Eq.~(\ref{eq:model}) is written for the first quadrant but easily translates for other three with appropriate
change of signs before $x$ and $y$. Fig.~\ref{fig:phot_model} schematically shows how the intensity is calculated in
such a model. The profile below the ray is identical to the pure Sersic brightness distribution. The brightness above the
ray decreases compared to values given by the Sersic's law as we move away from the ray. We experimented
with different laws of density declining. For example, we also considered a Gaussian type of declining but found that
the Guassian type and the exponential type produces nearly the same results. We have chosen the exponential law as one
of the simplest. We also note the horizontal direction of the density declining in Eq.~(\ref{eq:model}) is the
result of some trial and error. We also considered density declining along the elliptical isophotes and a straight
line perpendicular to the ray of the X-structure. In both of these cases, we found that such models produce some
unrealistic X-shaped bulges when applied to our sample of real galaxies.
\par
One of the important parameters that our model inherits from the Sersic function is an ellipticity of the isophotes
$\epsilon$. But in contrast to the simple Sersic model, the ellipticity parameter in this new model characterises only
that part of the isophotes that is below the ray of X-structure, while the shape of isophotes above the ray is
determined both by ellipticity and $s_\mathrm{X}$ parameter. For technical reasons, we define the ellipticity in the
following way:
\begin{eqnarray}
\epsilon = 
\begin{cases}
1 - b/a, \, b \leq a \\
a/b - 1, \, b > a,
\end{cases}
\label{eq:ell}
\end{eqnarray}
where $a$ and $b$ are the major and minor axes lengths of the given isophote, respectively.  According
to Eq.~(\ref{eq:ell}), $\epsilon$ can take values from $-1$ to $1$, or, in other words, we allow the model to have prolate
isophotes between the disc plane and the rays of X-structure.
\par
The photometric model from~\cite{Savchenko_etal2017} is almost identical to our photometric model~(\ref{eq:model}) in
terms of the statistic value per pixel and even slightly better, 0.7 versus 0.8, respectively. Although both of them are
a significant improvement over the Sersic model with a statistic value of 1.8. The residue images also look similar
(Fig.~\ref{fig:phot_model}). To our surprise, the residues show clear asymmetries at the level about 10\% of the total
intensity (see colorbar annotation). Probably, such asymmetries are remnants of the buckling event in this particular
galaxy model. It is also interesting that the model from~\cite{Savchenko_etal2017} applied to the pure bar does not show
the dark blobs between the rays of X-structure that are observed in almost all of \cite{Savchenko_etal2017}'s
X-structure images for real galaxies. Perhaps this means that the blobs are not due to the bulge model, but due to their
disc model.
\par
Our new photometric model is characterised by nine different parameters in total. Four of them are common parameters of
any component, such as the coordinates of the centre, the positional angle of the component, and the central
intensity. The other three are inherited from the Sersic model. These are the effective radius of the Sersic profile
$r_\mathrm{X}$, the Sersic index $n_\mathrm{X}$ and the ellipticity of the isophotes below the rays
$\epsilon_\mathrm{X}$. Here we have added a subscript to avoid further confusion with a simple Sersic model. The other
two parameters are specific for this particular model and include the opening angle of the X-structure ray $\varphi$,
and the scale length $s_\mathrm{X}$, which characterises the density dip above the ray. In practise, we found that the
ratio $r_\mathrm{X}/s_\mathrm{X}$ is a good indicator of how pronounced the X-structure we
observe. In Fig.~\ref{fig:comp_sharp} we give several examples of the model appearance for different values of this
parameter. It can be seen that the lower its value, the smaller the density dip between the rays, and the closer our
model is to the usual Sersic model. For shortness, we call this parameter the sharpness parameter in further
discussion. In principle, one can associate the value of this parameter with the shape of the B/PS bulge to indicate if
it has a rather boxy shape or peanut-like shape. However, the ellipticity $\epsilon_\mathrm{X}$ also affects the
intensity profile above the ray. This is due to the fact that we truncate the intensity of a Sersic photometric model
that has elliptical isophotes on its own. In Fig.~\ref{fig:comp_ell} we show how the model changes with ellipticity
$\epsilon_\mathrm{X}$ for a fixed value of $r_\mathrm{X}/s_\mathrm{X}$. It is clearly seen that the overall shape of the
profile becomes more boxy-like with a decrease in $\epsilon_\mathrm{X}$. At the same time, we observe a prominent
X-structure in all cases. The figure shows that the degree of boxyness can be strictly compared only among the models
with similar values of $r_\mathrm{X}/s_\mathrm{X}$. The pleasant side of a such rather complex behaviour is that the
model can account for cases when we observe a boxy-shaped bulge but with a prominent X-shape on top of
it.  
\par
Concluding this section, we obtained the photometric model of the B/PS bulge, at which the X-structure opening angle
value is set explicitly. Since the model now has four explicit rays, it is more accurate to call it ``the X-shaped bulge
model'' (or XB model for short) rather than the B/PS bulge model. To be definite, we will refer to our model in this
way.


\section{The fitting strategy}
\label{sec:fit}
Since our photometric model has two more parameters than the usual Sersic bulge model the fitting of the model requires
some additional care. In this section, we describe how we approached this problem. First of all, we assume that the
Sersic index $n_\mathrm{X}$ for the X-shaped bulge should be close to one. This assumption is based on the fact that the
bars were once part of the discs in which they reside and usually have a density profile of exponential type. Therefore,
we work with a fixed value of the Sersic index $n_\mathrm{X}=1$ for both real and modelled galaxies. For safety, we
performed additional runs with a free value of $n_\mathrm{X}$ for some of the simulated galaxies and found that it is
indeed close to one for various models (from about 0.9 to 1.3, more precisely).
\par
The fitting itself was performed in an iterative manner to let us move from a simple initial to a complex final model.
We divided the whole procedure into three stages, namely:
\begin{enumerate}
\item preliminary run; the purpose of this run is to determine the appropriate initial conditions for the X-shaped bulge model.
\item auxiliary run; this run softens the transition between the preliminary run and the final runs.
\item final run;  the final run consists of the actual fitting of the X-shaped bulge model along with other photometric components. 
\end{enumerate}  
The details of the runs are as follows. During the preliminary run we find the best-fit photometric model of the galaxy by
fitting B/PS bulge by the pure Sersic function with a fixed Sersic index value, $n=1$, and a fixed value of the positional
angle (PA), $\varphi_\mathrm{PA}=90^\circ$. At this stage, we do not use our X-shaped bulge model yet. The fitting is
done along with other components such as disc, ring, etc. After we obtained the optimal parameters of the pure Sersic
bulge, we construct the initial conditions for the X-shaped bulge model based on the pure Sersic bulge parameters. Here we
should choose values for only two additional parameters, $\varphi$ and $s_\mathrm{X}$. All others can be directly taken
from the previously obtained pure Sersic model. The characteristic range of the opening angles of real galaxies is from
20$^\circ$ to 40$^\circ$. The value of the opening angle can be chosen pretty safely near $30^\circ$ then. The scale
length $s_\mathrm{X}$ can be estimated in different ways but we choose to equate it to the characteristic vertical
scale length of the disc component $z_\mathrm{d}$. When we obtained the initial set of the parameters for the X-shaped
bulge photometric model, we again perform the photometric fitting of all components from the preliminary run now with
the X-shaped bulge model. However, we do not free all the parameters of the X-shaped bulge at once. We fist perform the
auxiliary run fitting the X-shaped bulge with a fixed value of PA, $\varphi_\mathrm{PA}=90^\circ$, and fixed coordinates
of its centre. At this step, we can also introduce some additional components such as a ring or a small Gaussian. Finally, we
free all the parameters of the X-shaped bulge model except for the Sersic index $n_\mathrm{X}$, which remains equal to
one till the end, and again perform the fitting. This is the final run.
\par
Another specific aspect of our research is that we consider a rather unusual disc model. We consider a model of the so-called anti-truncated disc (AT disc hereinafter) which is characterised by the presence of the hole in its centre:
\begin{equation}
j(r,z)= j_0 \sech^{2/n}\left(\frac{nz}{2z_0}\right) \times
\begin{cases}
  \exp\left(-\displaystyle\frac{r}{h}\right) , \; r \geq R_\mathrm{T}, \\
  \exp\left(-\displaystyle\frac{r}{h}-\displaystyle\frac{R_\mathrm{T}-r}{h_\mathrm{T}}\right),\; r < R_\mathrm{T}. 
\end{cases}
\label{eq:disc_model}
\end{equation}
Here $j_0$ is the central luminosity density, $h$ is the exponential scale length in the disc plane, $z_0$ is the vertical scale height, $n$ determines to which type of profile the vertical distribution is closer, $\sech^2$ with $n=1$ or $\exp(-z/z_\mathrm{d})$ with n$\rightarrow \infty$, $R_\mathrm{T}$ is the size of the hole and $h_\mathrm{T}$ is the characteristic scale length of the density decrease towards the centre. Eq.~(\ref{eq:disc_model}) is written in a cylindrical coordinate system  $(r,z)$. The actual intensity in the given pixel is obtained by integrating $j(r,z)$ along a LoS. 
\par
The reasons to consider AT discs in this study are the following. In general, AT discs are frequently discussed in
context of bars although not all bars reside in such discs and not all AT discs owe their existence to bars~(see, for
example,~\cite{Kim_etal14} and references therein)\footnote{We note that the observational studies usually consider
  discs with breaks as in~\cite{Kim_etal14} rather than AT discs. For further discussion, we assume that these two
  photometric models, although quite different, are used to describe the same physical situation at least in the context
  of SB galaxies, namely the change of density profile slope due to the bar.}.  However, the discs in our models indeed
have apparent AT profile if seen face-on (see Fig.~\ref{fig:bar}). The problem then if we decompose a galaxy applying a
simple exponential disc model while the disc actually has a hole in the centre we distort B/PS bulge profile in some
non-trivial way. If we then apply AT disc model the distortion effect should be compensated at least partially (at least
it is reasonable to assume so). We discuss the matter of different disc models more thoroughly in the next subsection
considering the decomposition of a particular numerical model.  The fitting of the AT disc model is not that simple
itself. In practice, we found that there are some cases when it is necessary to assign some reasonable values to the
hole parameters $R_\mathrm{T}$ and $h_\mathrm{T}$ and fix them until we find a rough approximation of the X-shaped bulge
model in the preliminary run. We free them only in the final run.
\par
Other photometric components we use are quite typical for studies of
this kind. We use a Gaussian ring, luminosity density of which is determined by the following expression:
\begin{equation}
j(r,z) = j(r_\mathrm{ring},0) \exp\left(-\frac{(r - r_\mathrm{ring})^2}{2 \sigma_\mathrm{ring}^2}\right) \exp\left(-\frac{|z|}{h_z}\right),
\end{equation}
where $r_\mathrm{ring}$ is the ring radius, $\sigma_\mathrm{ring}$ is the ring width, $h_z$ is the scale height and $j(r_\mathrm{ring},0)$ is the luminosity density at $r=r_\mathrm{ring}$. In some cases we also use simple Gaussian:
\begin{equation}
I(a)=I_0 \exp(-a^2/(2\sigma^2)).
\end{equation}
Here $\sigma$ is the characteristic size and $a$ is the distance between the given pixel and the Gaussian centre.  To
account for the dust in real galaxies we use the same method as in~\cite{Savchenko_etal2017}. We represent the dust
component by an edge-on disc with the negative intensity. The edge-on disc intensity is calculated using the following
expression:
\begin{equation}
I(r,z) = \mu(0,0) \frac{r}{h} K_\mathrm{1}\left(\frac{r}{h}\right) \sech^{2/n}\left(\frac{nz}{2z_0}\right),
\end{equation}
where $\mu(0,0)$ is the central surface brightness, $K_\mathrm{1}$ is the modified Bessel function of the first kind and all other notations are the same as in Eq.~(\ref{eq:disc_model}).
\par
To estimate uncertainties of the parameters we use bootstrap resampling available in {\small IMFIT} package. We
performed 200 iterations for each of the fits as recommended by the {\small IMFIT} author.
\par
Below we present results obtained from the decomposition of models and real galaxies. We start with numerical models and
show how various effects that are intrinsically non-physical (like projection effects, which include both possible disc
inclination and the rotation of the bar around the disc axis) and the effects that associated with the choice of
photometric model (in particular, the disc model) affect the output parameters of B/PS bulges. We also investigate the
behaviour of a particular model with a classical bulge to check how the presence of a bulge of another physical nature
affect the measured B/PS bulge parameters. Then we compare the parameters obtained from the physically different models
and finally compare the latter with those obtained from the decomposition of real galaxies.
\section{Numerical models}
\label{sec:numerical_models}

\begin{table*}
\centering
\caption{Parameters of models}
\begin{tabular}{ c | c | c | c | c | c | c | c | c }
\hline
$M_\mathrm{h}(R < 4R_\mathrm{d})/M_\mathrm{d}$ & $z_\mathrm{d} / R_\mathrm{d}$ & $Q (2R_\mathrm{d})$ & $M_\mathrm{b}$ & $r_\mathrm{b}$ &$N_\mathrm{d}$, $10^6$ &$N_\mathrm{h}$, $10^6$ &$\varepsilon_\mathrm{d}/R_\mathrm{d}, 10^{-3}$&$\varepsilon_\mathrm{h}/R_\mathrm{d}, 10^{-3}$ \\
\hline 
 1.5 & 0.05  &  1.2 &0& - & 4 &  4.5 & 3.68 & 12.8 \\
 \hline
 1.5 & 0.05  &  1.2 & 0.2 & 0.2 & 4 &  4.5 & 3.68 & 12.8\\
\hline
 1.5 & 0.05  &  1.6 &0& - & 4 &  4.5 & 3.68 & 12.8 \\
 \hline
 3.0 & 0.05  &  1.2 &0& - & 4 &  9  & 3.68 & 10.2 \\
 
\hline

\multicolumn{9}{p{0.8\textwidth}}
{\footnotesize{\textit{Notes}: each column represents parameters of the models, one model per line. $M_\mathrm{h}(R < 4R_\mathrm{d})$ is the mass of the halo within a sphere with a radius of $4R_\mathrm{d}$, where $R_\mathrm{d}$ is the scale length of the disc, $M_\mathrm{d}$ is the total disc mass, $z_\mathrm{d}/R_\mathrm{d}$ is the initial ratio of the disc scale height to the disc scale length, $Q(2R_\mathrm{d})$ is the value of the Toomre parameter at $2R_\mathrm{d}$, $M_\mathrm{b}$ and $r_\mathrm{b}$ are the total mass and the scale length of the bulge, respectively. $N_\mathrm{d}$ and $N_\mathrm{h}$ are the numbers of particles in the disc and the halo, respectively, and $\varepsilon_\mathrm{d}$ and $\varepsilon_\mathrm{h}$ are the softening lengths of the disc and halo particles, respectively.}}
\end{tabular}
\label{tab:models_pars}
\end{table*} 

In the present work we consider four different galaxy models from~\cite{Smirnov_Sotnikova2018}. Full details of the
simulations can be found in that work and we refer an interested reader to it. Here we briefly describe some important
aspects of the simulations. Initially, each model consists of an axisymmetric exponential disc
\begin{equation}
  \rho(R,z) = \frac{M_\mathrm{d}}{4\pi R_\mathrm{d}^2 z_\mathrm{d}} \exp\left(-\frac{R}{R_\mathrm{d}}\right)
  \sech^2\left(\frac{z}{z_\mathrm{d}}\right) \,,
\label{eq:sigma_disc}
\end{equation}
where $M_\mathrm{d}$ is the total mass of the disc, $R_\mathrm{d}$ is the disc scale length and $z_\mathrm{d}$ is the disc scale height, plus spherically symmetric dark halo of NFW-type~\citep{NFW}. The halo is characterised \textcolor{black}{by the ratio $M_\mathrm{h}(r<4R_\mathrm{d})/M_\mathrm{d}$}, where $M_\mathrm{h}(r<4R_\mathrm{d})$ is the dark halo mass inside the sphere with a radius of $4R_\mathrm{d}$. One model also includes a Hernquist type bulge~\citep{Hernquist1990}:
\begin{equation}
\rho_\mathrm{b}(r) = \frac{M_\mathrm{b}\, r_\mathrm{b}}{2\pi\,r\,(r_\mathrm{b} + r)^3} \,,
\end{equation} 
where $r_\mathrm{b}$ is the scale parameter and $M_\mathrm{b}$ is the total bulge mass. Disc and halo components are
represented by several millions of particles (see Tab.~\ref{tab:models_pars}). Bulge particle number is chosen such that
the mass of one particle from the bulge is equal to the mass of one particle from the disc. Some important physical
parameters of the models are specified in Tab.~\ref{tab:models_pars}. The initial equilibrium state of each
multi-component model was prepared via script {\tt{mkgalaxy}} \citep{McMillan_Dehnen2007} from the toolbox for N-body
simulation {\tt{NEMO}} \citep{Teuben_1995}. Evolution of the models was followed for about $8$ Gyr using the fastest
$N$-body code for one CPU {\tt{gyrfalcON}} \citep{Dehnen2002}. Each model gives rise to a bar after $1-2$ Gyr from the
beginning of the simulation. Each bar gradually thickens in the vertical direction (clear buckling events can
occur depending on the model) and takes a B/PS shape if seen side-on. Fig.~\ref{fig:bar} shows the
end-state of the bulgeless model with $\mu=1.5$ and $Q(2R_\mathrm{d})=1.2$. This model is our fiducial or ``main''
model. On the example of this model, we consider below how various projection effects affect the extracted parameters
of the B/PS bulge. For each of the models, we prepared a set of FITS images obtained from the corresponding density
distribution of disc/bulge particles. Such images can then be decomposed using a usual decomposition procedure analogous
to that used for images of real galaxies.
\subsection{Results for numerical models}

Most of the results presented in this subsection concern various non-physical effects like projection effects or the impact
of different photometric models of the disc on the X-shaped bulge parameters. To analyse such effects, we use our
``main'' model and choose a specific time moment, namely $T {\approx} 6 $ Gyr after the beginning of the simulation. To
this moment, the buckling events have already passed and the B/PS bulge evolves only slowly, with almost no changes in
the X-structure morphology. Using the stellar component density distribution for this moment, we prepared a set of FITS
images for \textcolor{black}{various orientations of the bar with respect to the disc axis (bar azimuthal angle)} and various inclinations of the disc plane. \textcolor{black}{The bar angle is calculated from LoS, that is, the angle is $0^\circ$ when the bar major axis is aligned along LoS and $90^\circ$ when the bar major axis is orthogonal to LoS.} 
\par
While our X-shaped bulge photometric model in principle is characterised by nine different parameters only four of them
are of interest for comparison with real observational data. The fact is that the value of any dimensional quantity (size,
intensity) measured from models depends on our normalisation procedure. To rightfully compare the dimensional parameters
with the observational data we need to know some quantities like the disc scale length when it was axisymmetric several
Gyrs before and the total disc mass. These characteristics are hard to obtain and they are subject to many errors. That
leaves us with four dimensionless quantities: the opening angle of X-structures rays $\varphi$, the sharpness parameter
$r_\mathrm{X}/s_\mathrm{X}$, the ellipticity of the isophotes $\epsilon_\mathrm{X}$ and Sersic index
$n_\mathrm{X}$. Here we present the results only for two of them, $\varphi$ and $r_\mathrm{X}/s_\mathrm{X}$. For
$n_\mathrm{X}$, we assume that it is equal to one in the present work. For ellipticity $\epsilon_\mathrm{X}$, we found
that it mainly falls in the range between $0.2-0.5$ for the considered galaxy models although there are some interesting
exceptions when it takes negative values. It also shows some trends depending on the bar viewing angle and the setup of the
photometric model. But, mostly due to lack of space and the desire to not overcomplicate the discussion, we prefer
to not discuss it here.

\subsubsection{Concept of "ideal" values of parameters} \label{sec:ideal}
Since numerical models allow one to extract a
bar by means of some dynamical considerations, we would like to exploit this possibility as much as possible to
understand more accurately the impact of various non-physical effects.
To this aim, we construct images depicting only bar particles like that shown in Fig.~\ref{fig:bar} for most of the
images of the simulated galaxies that we decompose below. We use such images with only a bar component to compare the
parameter values of the B/PS bulges obtained from the usual multi-component decomposition with the parameter values
obtained if there were no disc screening the B/PS bulge.
 Essentially, analysis of the ``pure'' B/PS bulges allows one to obtain ``ideal" values of parameters in a sense that
 parameters are then independent from the chosen disc model and the models of other photometric components. Comparison
 with ``ideal'' values of parameters allows us to understand more clearly how changes in the photometric model (for
 example, in disc model) affect the results of the decomposition.

\subsubsection{Different photometric models of the disc}
\begin{figure}
\begin{minipage}[t]{0.45 \textwidth}
\includegraphics[scale=0.5]{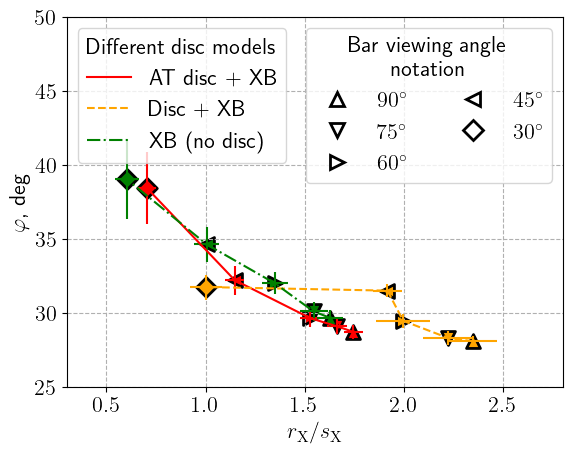}
\caption{Impact of different photometric models of the disc and \textcolor{black}{the bar viewing angle} on the measured values of opening angle $\varphi$ and the sharpness parameter $r_\mathrm{X}/s_\mathrm{X}$ of the X-shaped bulge photometric model.}
\label{fig:compare_discs}
\end{minipage}
\end{figure}

In photometric studies of the disc galaxies, it is customary to consider the photometric model of an exponential disc and its
derivatives (broken exponential, truncated, anti-truncated discs). Here we investigate how the choice of the photometric
model affects the parameters of B/PS bulge. We consider two types of disc models: a simple exponential disc and an AT
exponential disc (see Section~\ref{sec:fit}). Fig.~\ref{fig:compare_discs} shows how the value of the opening angle
$\varphi$ and the sharpness parameter $r_\mathrm{X}/s_\mathrm{X}$ depend \textcolor{black}{on the bar azimuthal viewing angle} for these two photometric
models and for a pure B/PS bulge (see the previous subsection). The figure contains some important information that
concerns the interpretation of observational data. First, if the bar is seen nearly side-on (azimuthal angle from
75$^\circ$ to 90$^\circ$), we observe a B/PS bulge with a quite prominent X-structure ($r_\mathrm{X}/s_\mathrm{X}$ from
1.5 to 2.5). In terms of opening angles, the choice of the disc model is rather insignificant in this case: the
difference between AT disc model and a simple exponential disc model is about $1^\circ$. Both of these models give
values of the opening angle that are smaller than ``ideal'' values (green curve) by about $2^\circ$, although they start to
diverge for smaller bar \textcolor{black}{rotation angles}. Thus, different disc models produce nearly consistent values of the opening angle for the
bar viewed side-on and the resulted values are close to ``ideal'' values in this case. However, in terms of another
parameter - the sharpness parameter - the application of different disc models lead to quite distinct measured values. An
application of a simple disc model leads to larger values of this parameter, i.e. it leads to a more pronounced
X-shape than it actually is.

\par
When we rotate the bar for more than about 30$^\circ$ \textcolor{black}{(up to $60^\circ$)} the B/PS bulge
shrinks to such a degree that the observed profile is substantially affected by choice of the disc model. Here we
observe a large discrepancy not only for the sharpness parameters but for the opening angles too. A simple exponential
disc model gives the values that are less than ``ideal'' ones by about
5$^\circ$-7$^\circ$.
 \par
 One more important thing we would like to note is that an AT disc model works a bit tricky for bars rotated less than 60$^\circ$. \textcolor{black}{For such bar orientations}, the model tends to increase its density scale length of the hole
 $h_\mathrm{T}$. In other words, the hole is gradually washed out as we rotate the bar. This happens because we observe
 the galaxies edge-on and observe the integrated luminosity profile. The problem is that the disc can have some
 additional asymmetric components like ansae (see Fig.~\ref{fig:bar}). With the bar rotation, they also rotate, and at
 some point we stop to distinguish the intensity dip between them. In practise, we fix $h_\mathrm{T}$ value if the ratio
 $R_\mathrm{T}/h_\mathrm{T}$ becomes greater than two. In this case, the model has a plateau in the centre instead of a
 hole. Nevertheless, Fig.~\ref{fig:compare_discs} indicates that such a degenerated AT disc model is still better than a
 simple exponential disc model, especially if the X-shaped bulge model turns out to have a small value of the sharpness
 parameter.  

\subsubsection{Impact of disc inclination}

\begin{figure*}
\begin{minipage}[t]{0.45 \textwidth}
\includegraphics[scale=0.5]{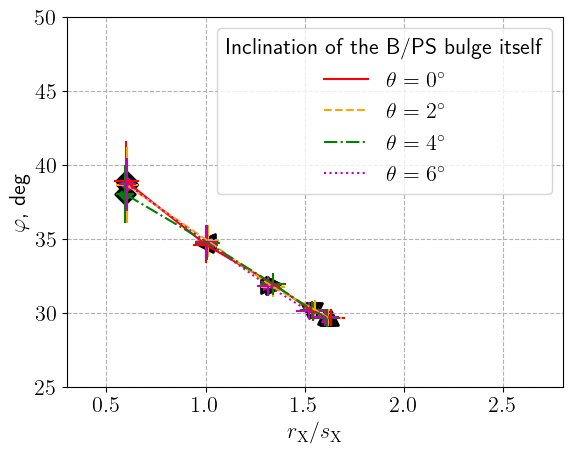}
\end{minipage}%
\begin{minipage}[t]{0.45 \textwidth}
\includegraphics[scale=0.5]{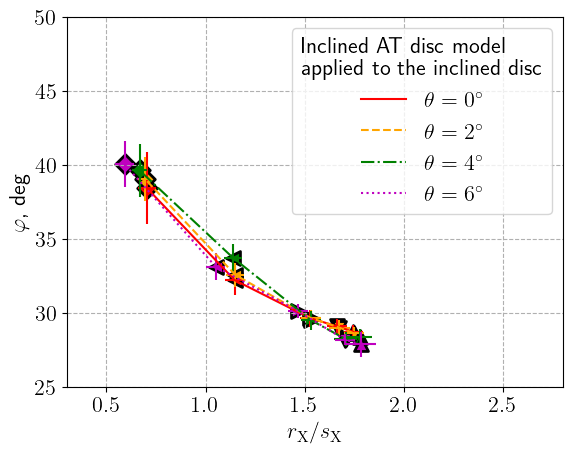}
\end{minipage}
\begin{minipage}[t]{0.45 \textwidth}
\includegraphics[scale=0.5]{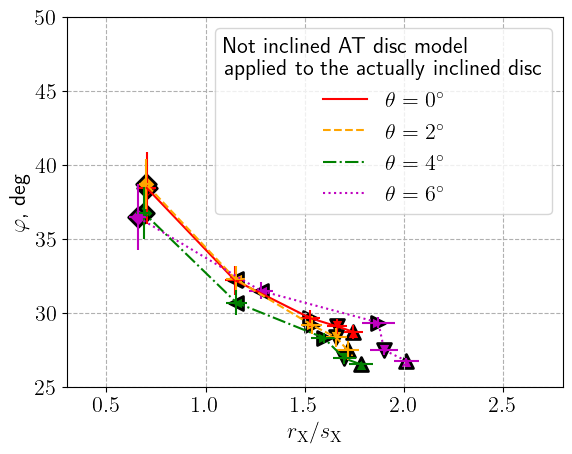}
\end{minipage}
\begin{minipage}[t]{0.45 \textwidth}
\includegraphics[scale=0.5]{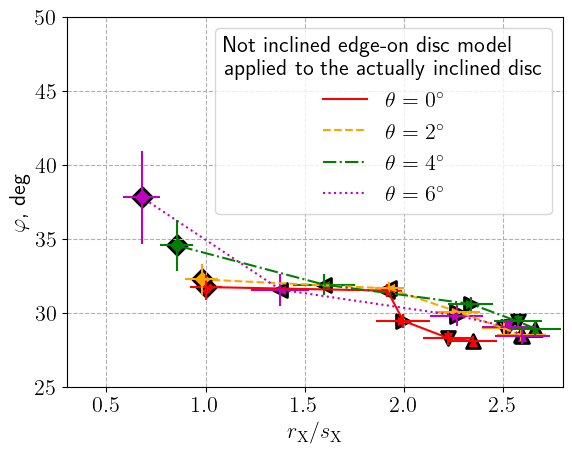}
\end{minipage}
\caption{Impact of different photometric models of the disc \textcolor{black}{and the bar viewing angle} (symbolic notation follows that from Fig.~\ref{fig:compare_discs}) on the the measured values of opening angle $\varphi$ and the sharpness parameter $r_\mathrm{X}/s_\mathrm{X}$ of the X-shaped bulge photometric model for different inclinations of the disc plane.}
\label{fig:compare_proj}
\end{figure*}

Although our sample of real galaxies mostly contains galaxies seen edge-on we should confirm whether our results depend
on small inclinations of the disc plane or not. A problem here is that the light coming from the inclined disc is mixed
with that produced by a B/PS bulge, and the more inclined disc we take the more the disc model affects the bulge. To
accurately analyse consequences of disc inclination, we considered four different cases. First, we check how the
parameters of B/PS bulge itself change with an inclination if there were no disc. To this aim, we take a pure B/PS bulge
extracted by means of the dynamical decomposition and constructed corresponding images for different values (from
2$^\circ$ to 6$^\circ$) of the disc inclination angle. Then we fit our X-shaped bulge photometric model to each of the
images. Fig.~\ref{fig:compare_proj} (\textit{top left}) shows the dependence of the opening angle $\varphi$ and the
sharpness parameter $r_\mathrm{X}/s_\mathrm{X}$ on disc inclination and \textcolor{black}{the bar azimuthal angle}. The answer that follows from the
figure is rather simple. A slightly inclined B/PS bulge is almost the same as not an inclined B/PS bulge. Both the
opening angles and the sharpness parameters have almost the same values for inclined and not inclined B/PS bulges. This
is an important result because the problem then has one less degree of freedom, which in principle could exist if the
properties of the bulge strongly depended on inclination.
\par
The next step is to consider how the presence of the inclined disc affects the parameters of B/PS bulges. At this step, we
analyse the whole model consisting of all stellar component particles. We check the behaviour of three different
photometric models of the disc: \textit{a)} inclined AT disc (Fig.~\ref{fig:comp_all}, \textit{top right}) \textit{b)}
not inclined AT disc applied to actually inclined disc (Fig.~\ref{fig:comp_all}, \textit{bottom left}) \textit{c)} not
inclined simple exponential disc applied to inclined disc (Fig.~\ref{fig:comp_all}, \textit{bottom right}). The last
two cases should also be considered, since the inclination of the disc, although it can be viewed as a free parameter,
cannot be reliably estimated in some cases due to the actual resolution of the image and the presence of other
photometric components. A general conclusion from the figures is that the disc inclination does change the values of the
opening angles and the sharpness parameter. The best photometric disc model that gives the values that are most
consistent with the ideal ones is an AT inclined disc model. In this case, there is almost no distinction between the
curves as in the case of the pure B/PS bulge. A photometric model of a not inclined AT disc but with the fixed
inclination $i=90^\circ$ applied to actually inclined disc results in the opening angles that differ from ``ideal''
values for about $2^\circ-3^\circ$ depending on the bar orientation and disc inclination. The value of the sharpness parameter
also slightly changes. As can be seen from the figure, an application of a simple disc model (without an inner
truncation) leads to a poor estimation of B/PS bulge parameters especially in case of a bar major axis close to LoS. The
situation repeats that for not inclined discs with no essential changes for the better or worse.
\subsubsection{Other photometric components. Case of co-existing bulges}
\begin{figure}
\begin{minipage}[t]{0.45 \textwidth}
\includegraphics[scale=0.5]{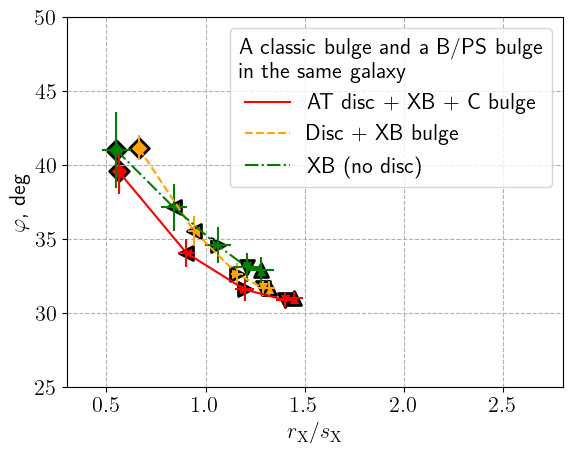}%
\end{minipage}%
\caption{The dependence of the opening angle $\varphi$ and the sharpness parameter $r_\mathrm{X}/s_\mathrm{X}$ of X-shaped bulge model on the \textcolor{black}{bar viewing angle} (symbolic notation follows that from Fig.~\ref{fig:compare_discs}) for three different combinations of dynamical subsystems (see text for details) of the galaxy model with co-existing classical bulge and B/PS bulge. ``C bulge'' stands for a classical bulge.}
\label{fig:compare_bulges}
\end{figure}
One of the possible problems of the photometric decomposition of disc galaxies is that the different types of bulges can
co-exist in one galaxy. A well-known example is NGC~4565 where two types of bulges are distinguished, a boxy bulge and
the so-called pseudobulge (see \cite{Kormendy_Barentine2010}). The co-existence of different bulge types were also
studied by~\cite{Erwin_etal15} where authors distinguished discy and classical bulges. Most of real galaxies from our
sample also seem to contain some central component which is not the part of the B/PS bulge. Therefore, we would like to
check how the parameters of our B/PS bulge photometric model are affected by the presence of a bulge of a different
physical nature. To this aim, we use a galaxy model from \cite{Smirnov_Sotnikova2018} that initially has a spherical
Hernquist bulge on top of the axisymmetric exponential disc (see Tab.~\ref{tab:models_pars}). This model has the same
parameters as the "main" model except it has a classical bulge component with $M_\mathrm{b}/M_\mathrm{d}=0.2$ and
$r_\mathrm{b}=0.2 R_\mathrm{d}$.  We analysed it in the same way as the previous ones. First, we performed the dynamical
decomposition and cut a B/PS bulge component based on the frequency analysis. Then we fit our X-shaped bulge photometric
model to images with only a B/PS bulge to obtain "ideal" values of parameters (see section~\ref{sec:ideal}). Next, we
consider two cases: the first, when the image is constructed only from disc and B/PS bulge particles, and the second,
when the image includes all components, namely, a disc, a B/PS bulge, and a classical bulge. In the first case, we fit XB
model along with AT disc and in a second one we fit a Sersic bulge photometric model along with XB and AT disc
components. The former case should be considered because the resulted deviation from ideal values (if any) can arise due
to the disc component. Fig.~\ref{fig:compare_bulges} show how the geometric parameters of B/PS bulge vary in all three
cases depending on the bar orientation. As can be seen from the figure, the values of opening angles and the sharpness parameter
are slightly different in each case. The presence of a classical bulge slightly lowers the opening angles although the
difference is not that large, about $3^\circ$ if the bar is rotated for about 45$^\circ$ and smaller for other bar
orientations. In general, Fig.~\ref{fig:compare_bulges} shows that we indeed can hope to retrieve the values of B/PS bulge
parameters close to real ones if several bulges co-exist in the same galaxy.

\subsubsection{Impact of the physical parameters of the galaxies on the B/PS bulge parameters}
\begin{figure}
\begin{minipage}[t]{0.5 \textwidth}
\includegraphics[scale=0.5]{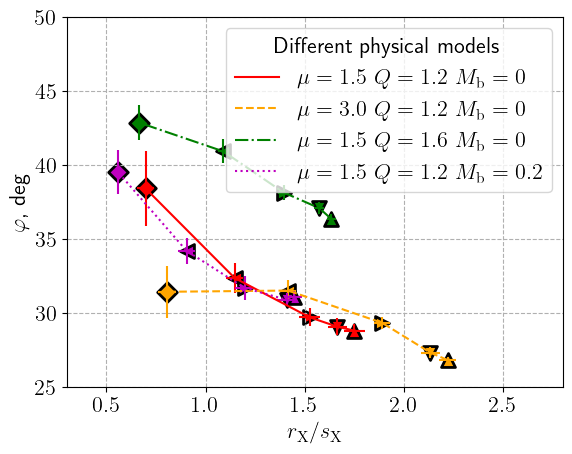}%
\end{minipage}%
\caption{The dependence of the opening angle $\varphi$ and the sharpness parameter $r_\mathrm{X}/s_\mathrm{X}$ of X-shaped bulge model on the bar \textcolor{black}{viewing angle} (symbolic notation follows that from Fig.~\ref{fig:compare_discs}) for four different galaxy models from Tab.~\ref{tab:models_pars}.}
\label{fig:compare_models}
\end{figure}
Different physical models of galaxies give rise to quite different types of bars (for example, bars with a barlens or a
peanut-shaped bars~\citep{Salo_Laurikainen2017, Smirnov_Sotnikova2018}). Naturally, the morphology of B/PS bulges that
are extensions of the bars in the vertical direction also depends on the galaxy physical parameters. General trends in
evolution of B/PS bulges and their X-structures in different disc models were thoroughly investigated
by~\cite{Smirnov_Sotnikova2018}. One of the important conclusions obtained by the authors was that the smallest opening
angles should be observed in galaxies where dark matter has a significant contribution to the overall gravitational
field within the optical radius of the disc, that is, in dark matter dominated galaxies. However,
in~\cite{Smirnov_Sotnikova2018} the angles were measured based on the density distribution along the photometric cuts.
First, we want to ensure that the same conclusion holds if we use our new photometric model of the X-shaped bulge. There
are also some other notable trends that were observed by~\cite{Smirnov_Sotnikova2018}. For example, opening angles of
the X-structure increase with an increase of the initial value of Toomre parameter $Q$ in the disc or with an addition of a
spherical bulge component to the model. We would like to verify these trends using our X-shaped bulge model. To this
aim, we consider four representative models from~\cite{Smirnov_Sotnikova2018} (see Tab.~\ref{tab:models_pars} in the
present work). For each of the models we select only one time moment again, namely $T {\approx} 6$ Gyrs after the
beginning of the simulations for models with $\mu=1.5$\footnote{This is the same one we work with in the previous
  subsections considering our ``main'' model.} and $T {\approx} 7$ for the model with $\mu=3$. We consider a later time
moment for the model with $\mu=3$ since the bar in this model is formed about 1 Gyr later than in other models. We
constructed images for each of the models for different bar orientations around disc axis and performed the photometric fitting in each case. All
images were fitted by the combination of AT disc and B/PS bulge photometric models. The results are presented in
Fig.~\ref{fig:compare_models}. As can be seen from the figure, the model with the heaviest dark halo ($\mu=3$) is indeed
distinct. It shows the smallest opening angle of all models (in case of a side-on bar), although the difference between
this model and our ``main'' model is not that large (${\approx}$ 2$^\circ$ for a side-on bar). At the same time, B/PS
bulges in these models are quite different in terms of the sharpness parameter. In the model with the heavy halo, we
observe a more prominent intensity dip between the rays (see Fig.~\ref{fig:comp_sharp} for reference).
We should note that the transition \textcolor{black}{of a bar azimuthal angle} from 45$^\circ$ to 60$^\circ$ in this model seems rather unnatural as the
opening angle value does not increase while the B/PS bulge substantially shrinks which is identified by a decrease in
the sharpness parameter value. Such a behaviour seem to be caused by a complex structure of the B/PS bulge itself. As
was shown by \cite{Parul_etal2020} the appearance of the X-structure ray can be associated with the existence of some
particular orbital family of quasi-periodic orbits. Thus, we can assume that the mix of equally populated orbital
families can produce the density peaks in a rather wide strip. But the overall picture should also depend on the bar \textcolor{black}{viewing angle}
since different families tend to occupy different regions of the bar~\citep{Portail_etal2015b,
  Parul_etal2020}. Therefore, if for a side-on bar we observe an imprint of only one orbital family there is a
possibility that we start to distinguish some other families along with the first one in case of a rotated bar. A
detailed orbital analysis of the model should clarify this issue.
\par
A model with $Q=1.6$ and a model with a classical bulge also show values of opening angles consisting with those
obtained by~\cite{Smirnov_Sotnikova2018}. The model with $Q=1.6$ shows a rather large opening angles for all bar \textcolor{black}{orientations} in
comparison the ``main'' model, from about 36$^\circ$ to 42$^\circ$.  The model with a classical bulge is close to the
``main'' model in terms of opening angles but show although not that distinct but still considerably smaller values of
the sharpness parameter.
In general, Fig.~\ref{fig:compare_models} demonstrates that the B/PS bulge parameters, as expected, depend both on the
physical conditions in the disc and the structure of other physical components. But what is important here is that using
our new photometric model we capture the different trends observed for different physical models like those obtained
by~\cite{Smirnov_Sotnikova2018}. It is also important that the sharpness parameter seems to be a good indicator of a bar \textcolor{black}{azimuthal viewing angle}. Although, we emphasise that we consider only four models here. Therefore, this statement should be strictly verified
on the larger sample of galaxy models.
\subsubsection{Finishing remarks about numerical models}
Concluding this subsection we would like to stress out that we do not pursue a goal to investigate as many possible
physical models as we can or trends in their evolution here. No, our goal here was different. We wanted to ensure that
the use of the X-shaped bulge photometric model leads to sensible results, that is:
\begin{enumerate}
	\item for different geometric configurations arising due to the possible disc inclination and the rotation of the bar major axis around disc axis we can hope to retrieve ``true'' values of the X-shaped bulge parameters,
	\item that we can do it in case of a composite inner structure of the galaxy (although we considered here a relatively simple case),
	\item the application of the X-shaped model at least reproduces some qualitative results obtained by~\cite{Smirnov_Sotnikova2018}.
\end{enumerate}
In a way, this can be called a ``debugging'' of the X-shaped bulge model, that is, we apply various tests to see whether
it works properly in different situations.
 \par
 It is also important that we obtained some reference points such as the lower end of the opening angles range which is
 about $26^\circ$. This value is measured in the model with $\mu=3$ for a side-on bar. In~\cite{Smirnov_Sotnikova2018}
 we showed that the X-structure in this model demonstrates the smallest opening angle of all thirteen models considered
 there. We expect that the B/PS bulges of real galaxies should rarely have X-structure angles smaller than this value,
 because to detect such a small value, one needs to detect a bar viewed side-on in a galaxy with a rather heavy
 dark halo, plus at later stages of its evolution. This is indeed possible, but should still be a relatively rare case.

\section{Real galaxies}
\label{sec:real_galaxies}
\subsection{Data}
In order to apply our method to the images of real galaxies, we compiled a sample of galaxies by combining the sample
of ~\cite{Ciambur_Graham2016} with the sample of \cite{Savchenko_etal2017}, that gave us in total 31 galaxies
with prominent B/PS bulges. Table~\ref{tab:sample} shows the general parameters of the sample galaxies compiled from
HyperLEDA \citep{2014A&A...570A..13M} and NASA/IPAC Extragalactic Database (NED)\footnote{\url{http://ned.ipac.caltech.edu/}}
databases. In this section we describe the data preparation for these galaxies.

Optical images were retrieved from the Sloan Digital Sky Survey DR12 database\footnote{\url{https://www.sdss.org/dr12/}}
\citep{2011AJ....142...72E} in r-band. The following data preparation pipeline was identical to the one from \cite{Savchenko_etal2017}
and included following steps for every image:
\begin{enumerate}
\item adjacent SDSS fields were combined into single image using {\small SWARP} package \citep{2002ASPC..281..228B};
\item background flux was approximated using 2D polynomial of objects free regions of the image and subtracted;
\item the position angle of the galaxy was estimated using method from \cite{2012MNRAS.427.1102M} and the image was
  rotated such that the galaxy was oriented along the x-axis of the image;
\item a point spread function of the image was constructed using a set of isolated non-saturated stars on the image;
\item all the objects on the image except for the galaxy were masked out using object catalogues made by {\small SEXTRACTOR}
  package \citep{1996A&AS..117..393B} to exclude them from the analysis.
\item \textcolor{black}{Sigma images were automatically generated using {\small IMFIT} software based on the gain, read noise, and intensity of the original subtracted sky background values which all can be found in SDSS database for the corresponding fields. The precise equation is the following:
\begin{equation}
\sigma_i^2= (I_{d,i}+I_\mathrm{sky})/g_\mathrm{eff}+N_\mathrm{c} \sigma_\mathrm{rdn}^2/g_\mathrm{eff},
\end{equation}  
where $I_{d,i}$ is the data intensity, $I_\mathrm{sky}$ is the intensity of the original subtracted sky background, $\sigma_\mathrm{rdn}$ is the read noise value, $N_\mathrm{c}$ is the number of separate images combined to form the data image, and $g_\mathrm{eff}$ is the ``effective'' gain (the product of the gain and $N_\mathrm{c}$).}
\end{enumerate}

The IRAC 3.6$\mu$m images were downloaded from the Spitzer Survey of Stellar Structure in Galaxies
database\footnote{\url{http://irsa.ipac.caltech.edu/data/SPITZER/S4G/}} (S4G, \cite{2010PASP..122.1397S}). These images were
also corrected for the background and rotated identically to the optical ones. Objects masks and sigma-images
were adopted from \cite{Salo_etal2015}, and we used instrumental PSF from the mission
website\footnote{\url{https://irsa.ipac.caltech.edu/data/SPITZER/docs/irac/calibrationfiles/psfprf/}}.
\textcolor{black}{For all SDSS and IRAC images the individual pixel weights $w_i$ were calculated as a reverse square of the corresponding sigma values, $w_i=1/\sigma_i^2$.}

\begin{table}
  \centering
  \begin{tabular}[h]{lccl}
    \hline
    Name         &   T  &  $M_r$ & $d_{25}$  \\
                 &      &   mag  &  arcmin  \\       
    \hline
    PGC002865    &  5.8 & -20.01 &   1.78   \\
    PGC010019    &  3.0 & -20.96 &   0.78   \\
    PGC021357    &  3.3 & -20.65 &   1.14   \\
    PGC024926    &  2.0 & -19.36 &   1.70   \\
    PGC026482    &  3.1 & -20.66 &   1.17   \\
    PGC028788    &  0.5 & -20.35 &   0.85   \\
    PGC028900    &  0.9 & -21.55 &   1.00   \\
    PGC030221    &  3.0 & -20.44 &   1.12   \\
    PGC032668    &  0.0 & -20.59 &   1.04   \\
    PGC034913    &  3.0 & -20.89 &   1.91   \\
    PGC037949    &  1.4 & -20.82 &   1.32   \\
    PGC039251    & -0.9 & -20.73 &   1.69   \\
    PGC044422    &  0.4 & -21.13 &   0.72*  \\
    PGC045214    &  1.0 & -21.41 &   0.89   \\
    PGC053812    &  3.9 & -21.51 &   1.10   \\
    PGC055959    &  2.0 & -21.38 &   1.02   \\
    PGC069401    &  1.0 & -21.97 &   1.10   \\
    PGC069739    &  5.2 & -20.92 &   2.00   \\
    ASK361026.0  &  4.6 & -21.77 &   0.44*  \\
    ESO443-042   &  3.0 & -20.90 &   2.81   \\
    NGC128       & -2.0 & -22.30 &   3.16   \\
    NGC678       &  3.0 & -20.98 &   3.09   \\
    NGC2549      & -2.0 & -20.07 &   3.63   \\
    NGC2654      &  2.0 & -19.69 &   4.47   \\
    NGC2683      &  3.0 & -19.82 &   9.55   \\
    NGC3628      &  3.1 & -19.49 &  10.96   \\
    NGC4111      & -1.3 & -19.68 &   1.78   \\
    NGC4469      &  0.3 & -19.07 &   2.88   \\
    NGC4710      & -0.9 & -20.53 &   4.37   \\
    NGC5529      &  5.1 & -20.71 &   5.75   \\
    NGC7332      & -1.9 & -19.33 &   2.95   \\
    \hline
  \end{tabular}
  \caption{Parameters of the sample galaxies: name, T-type, absolute magnitude in r-band and the apparent
    diameter.}
  \label{tab:sample}
\end{table}

\subsection{Results}
Although we assume that B/PS bulges of the modelled galaxies should resemble the bulges of real galaxies, the physics of
real galaxies is definitely richer. In particular, the influence of a gaseous component, star formation, supernova
feedback, not to mention our doubts as to whether we model the physics of dark matter correctly, are the main factors
that we miss in our models. Although these factors are not directly associated with B/PS bulges they determine the
properties of the bar (its pattern speed and size) thus indirectly affecting the B/PS bulges too. We also do not possess
knowledge of initial conditions in the disc when it began to form the bar. All these problems are fairly well-known
problems of numerical modelling which are currently addressed in many numerical studies. Therefore, if some deduced
photometric model works fairly well in the case of modelled galaxies it does not necessarily adequately represent B/PS
bulges in real galaxies. To clarify this, we start this section with a few important examples showing how the X-shaped
bulge model works in the case of real galaxies.

\subsubsection{NGC~128}
\begin{figure*}
\begin{minipage}[t]{0.5 \textwidth}
\includegraphics[scale=0.1]{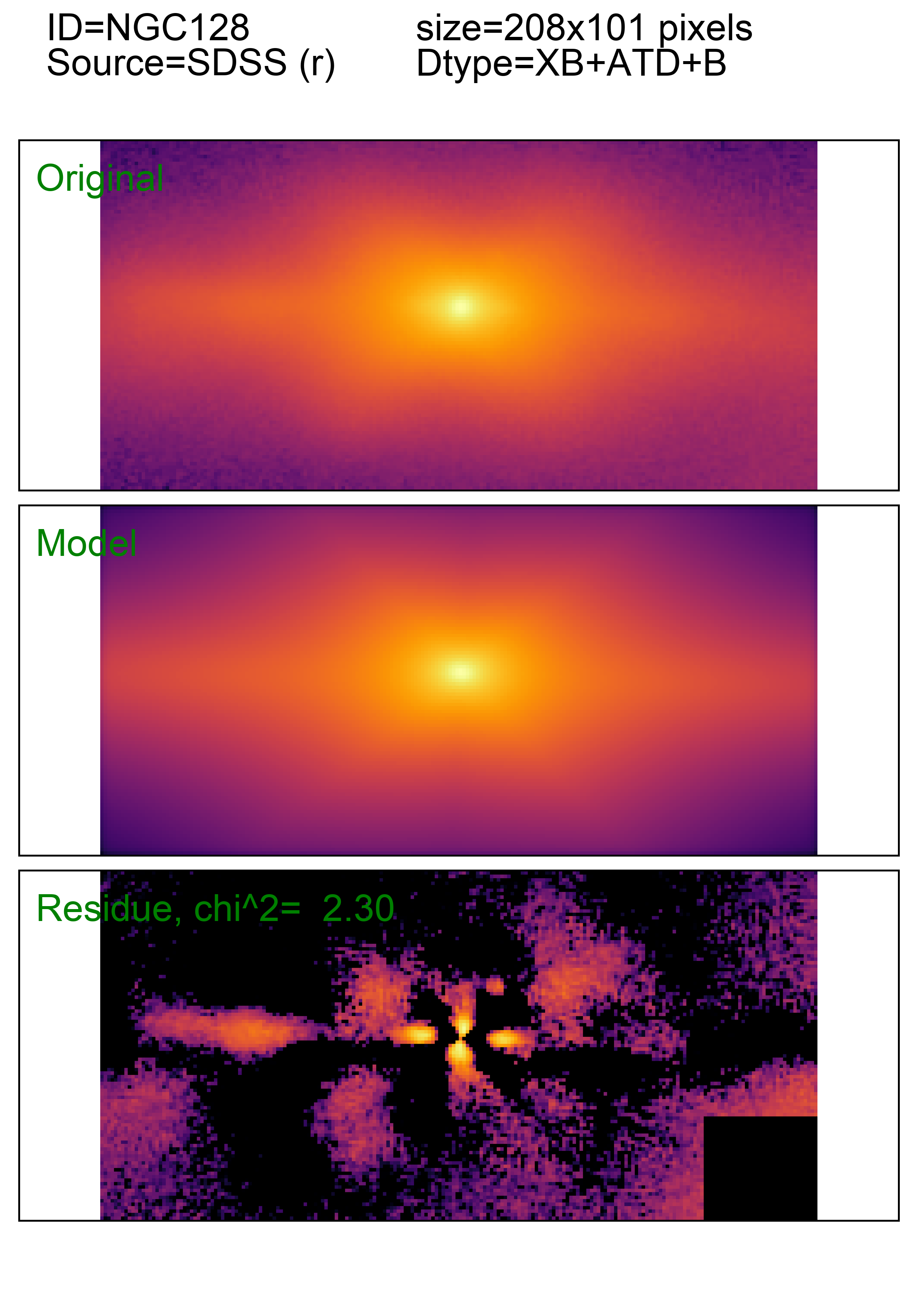}%
\end{minipage}%
\begin{minipage}[t]{0.5 \textwidth}
\includegraphics[scale=0.1]{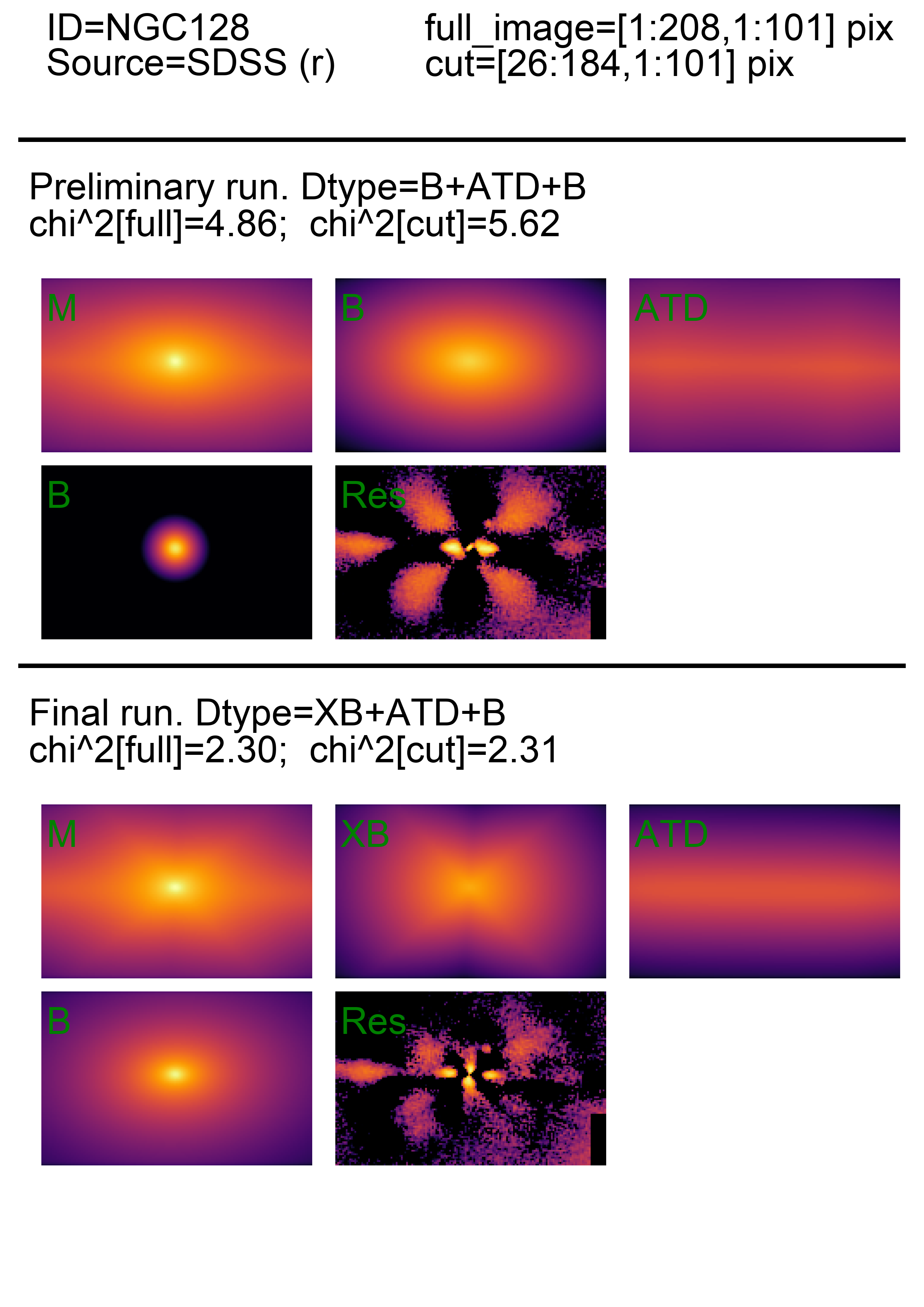}%
\end{minipage}%
\caption{The photometric decomposition of NGC~128. \textit{Left:} the original galaxy image (\textit{top}), the best-fit
  photometric model (\textit{middle}) and the residue image (\textit{bottom}). \textit{Right:} the images of the
  individual photometric components and the residue image for the preliminary run with a simple Sersic bulge
  (\textit{top}) and for the run with the X-shaped bulge model given by Eq.~(\ref{eq:model}) (\textit{bottom}). The
  auxiliary text above the images gives various details of the decomposition. We note that the images in the right
  column show the rectangle of a smaller area than full images from the left column (see auxiliary text). This smaller
  area roughly corresponds to the area where the most part of the B/PS bulge resides. We give two chi-square values, one
  of which is calculated over all pixels used for the decomposition ("chi\textasciicircum2[full]") and the second one
  which is calculated over the pixels only from this smaller area ("chi\textasciicircum2[cut]"). The latter is more
  sensitive to the B/PS bulge model and is more convenient to use if we want to compare different models of the B/PS
  bulge.}
\label{fig:NGC128}
\end{figure*}

  
We start with one of the most prominent candidates among the galaxies with a B/PS bulge and X-structures, NGC~128. This
galaxy is a famous galaxy where the first X-structure was distinguished \citep{Burbidge_Burbidge1959}. Together with
NGC~126 (E/SB0), NGC~127 (Sa), and NGC~130 (E5) and possibly NGC~125 (S0 pec) \citep{Zwicky_etal1965,
  DOnofrio_etal1999}, this galaxy forms a group of galaxies~\citep{Peterson1979}. The outer disc region of NGC~128 is
substantially disturbed by the gravitational interaction with NGC~127, which causes the appearance of a bridge
connecting one to another~\citep{Burbidge_Burbidge1959}. The analysis of the gas kinematics showed that the central area
($\sim$5 arcsec, roughly one-quarter of the peanut projected size) harbours the counter-rotating gaseous
component~\citep{DOnofrio_etal1999}. Nearly on the same space scale (about one fifth of the larger peanut
size)~\cite{Ciambur_Graham2016} identified the so-called ``nested peanut'' residing in the larger one. Needless to say,
this galaxy is not an easy target for the 2D photometric decomposition. Mostly due to its peculiar inner
structure.  \par
The decomposition procedure follows our usual pattern. It consists of two stages, namely a preliminary stage at which we
find the initial conditions for the X-shaped bulge fitting the B/PS bulge by a Sersic function and the final stage where
we start from previously obtained set of parameters to find the best-fit model. We found by trial and error that there
is a problem with the before-mentioned inner component boiling down to the fact that it should be somehow represented in
the initial photometric model. Otherwise, the outer X-shaped bulge tends to degenerate into the inner bulge. Therefore,
our initial photometric model consists of AT disc and two bulges, one of which has a fixed Sersic index $n=1$. The
second one has free parameters except for the PA which is fixed in order to prevent the bulge degeneration into two
lobes of the B/PS bulge at the preliminary run (we free this parameter at a final run). The results of our decomposition
procedure (best-fit model and residue images), as well as individual images of all components for two runs are shown in
Fig.~\ref{fig:NGC128}. We exclude the outer part of the disc from fitting because of the disturbance caused by the
gravitational interaction with NGC~127, and fit only the smaller area that nevertheless contains the whole peanut and
small ansae that can indicate where the disc inner truncation should start. Chi-square values (see the caption for the
individual components images) indicate that the modification of the Sersic bulge to the X-shaped bulge leads to a
significant improvement of the residue. Four lobes appearing in the preliminary run residue image are mostly dispersed
to the final run and the central peculiar component becomes the main source of errors. We also note that the second
bulge representing an inner component becomes close to the X-shaped bulge in terms of their linear scales
($r_\mathrm{X}/r_\mathrm{B}{\approx}1.2$). Although this issue is not that severe if we consider their respective Sersic
indexes which are quite different ($n_\mathrm{X}=1$ versus $n=2.6$).
\par 
We performed an additional run with a fixed to zero ellipticity of the inner component to see whether we can somehow
lessen the degeneracy. In other words, we try to fit a bulge component with completely circular isophotes. Thus, we
assume that the inner component is rather not a nested peanut but more like a small pseudobulge arising due to gas
concentration in the central area of the galaxy. Due to lack of space, we do not provide the corresponding figure
here, but it is available in online materials (see NGC~128* there). We found that for such a restricted model of an
inner component the X-shaped bulge and the Sersic bulge representing the inner component turn out to have quite distinct
sizes ($r_\mathrm{X}/r_\mathrm{b}{\approx}6$), which implies a lesser degree of degeneracy. However, the residue worsens
and the corresponding chi-square value increases from 2.3 to 3.0. Although, it is still better than a chi-square value
of a preliminary run, $\chi^2=5.62$.
 \par
 It should be stressed out that in both types of the decomposition (with the fixed and free ellipticity) the resulted
 values of opening angles and linear scales of the X-shaped bulges are close to each other ($\varphi=44^\circ$ in both
 cases and $r_\mathrm{X}\approx3.0$ kpc for the free run and $r_\mathrm{X}\approx2.7$ kpc for the run with the fixed
 ellipticity). There are indeed some notable changes in the value of the second length scale $s_\mathrm{X}$ of the
 X-shaped bulge, $s_\mathrm{X}\approx2.7$ kpc versus $s_\mathrm{X}\approx4$ kpc, respectively, and according to that the
 sharpness parameters have quite different values too ($r_\mathrm{X}/s_\mathrm{X}\approx1.11$ versus
 $r_\mathrm{X}/s_\mathrm{X}\approx0.67$). In other words, what changes between these two runs is the sharpness of the
 observed X-structure. In a sense, this behaviour resembles the behaviour of different disc models applied to the same
 numerical model which we investigated in Section~\ref{sec:numerical_models}. In that case, there were also changes in
 the sharpness parameter while the opening angles turn out to be close to each other.
\par
We should also note some important things concerning the nature of the inner component (nested peanut?). The residue
image of a typical peanut usually shows four pronounced lobes (see Fig.~\ref{fig:NGC128}, preliminary run) if we fit the
B/PS bulge by a Sersic function. If we assume that the inner component is a nested peanut similar to the outer one, then
it is reasonable to further assume that the residue image should show four inner lobes in this case. There are indeed
four lobes seen in the residue image (Fig.~\ref{fig:NGC128}, final run), but two of them appear in the disc plane and
the other two sit on the line perpendicular to the disc plane. Therefore, the residue of the inner structure does not
resemble the residue of the outer structure. Although there may still be some doubt, since the residue of the central
area was obtained by subtracting the B/PS bulge and disc photometric models, we believe that the structure of the
residue indirectly indicates that the inner component is at least not an ordinary peanut. In principle, such a structure
of the residue can appear due to the presence of the inner (second) small bar if its major axis is oriented close LoS
and the disc plane is slightly inclined. \textcolor{black}{Based only on the residue structure we also cannot rule out the possibility that there is simply a small flattened bulge (either classical of pseudo-bulge formed via gas inflow and subsequent star formation)}. We are not going to purse further this question here and leave it for
future studies.

\subsubsection{ESO~443-042}
\begin{figure*}
\begin{minipage}[t]{0.5 \textwidth}
\includegraphics[scale=0.1]{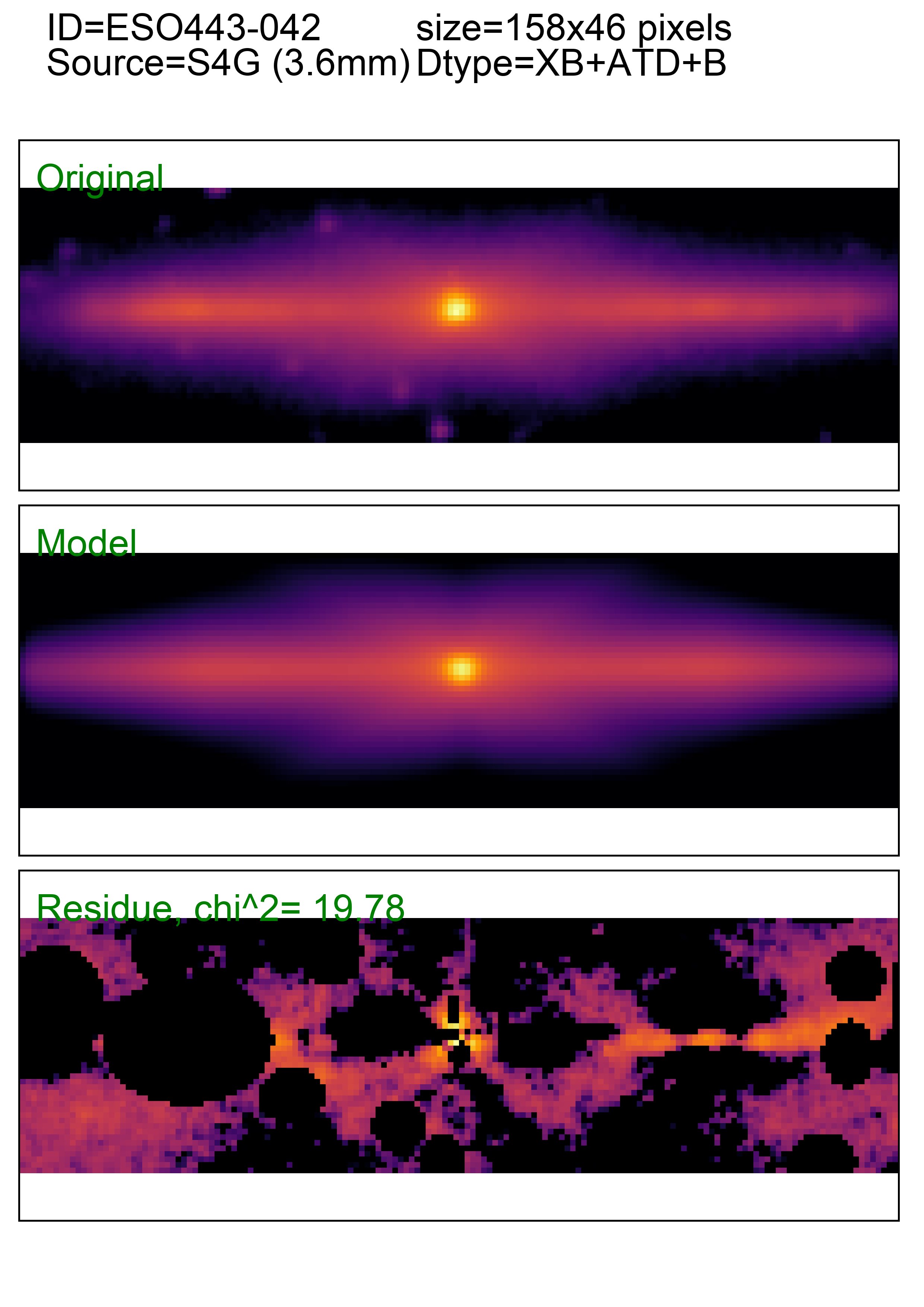}%
\end{minipage}%
\begin{minipage}[t]{0.5 \textwidth}
\includegraphics[scale=0.1]{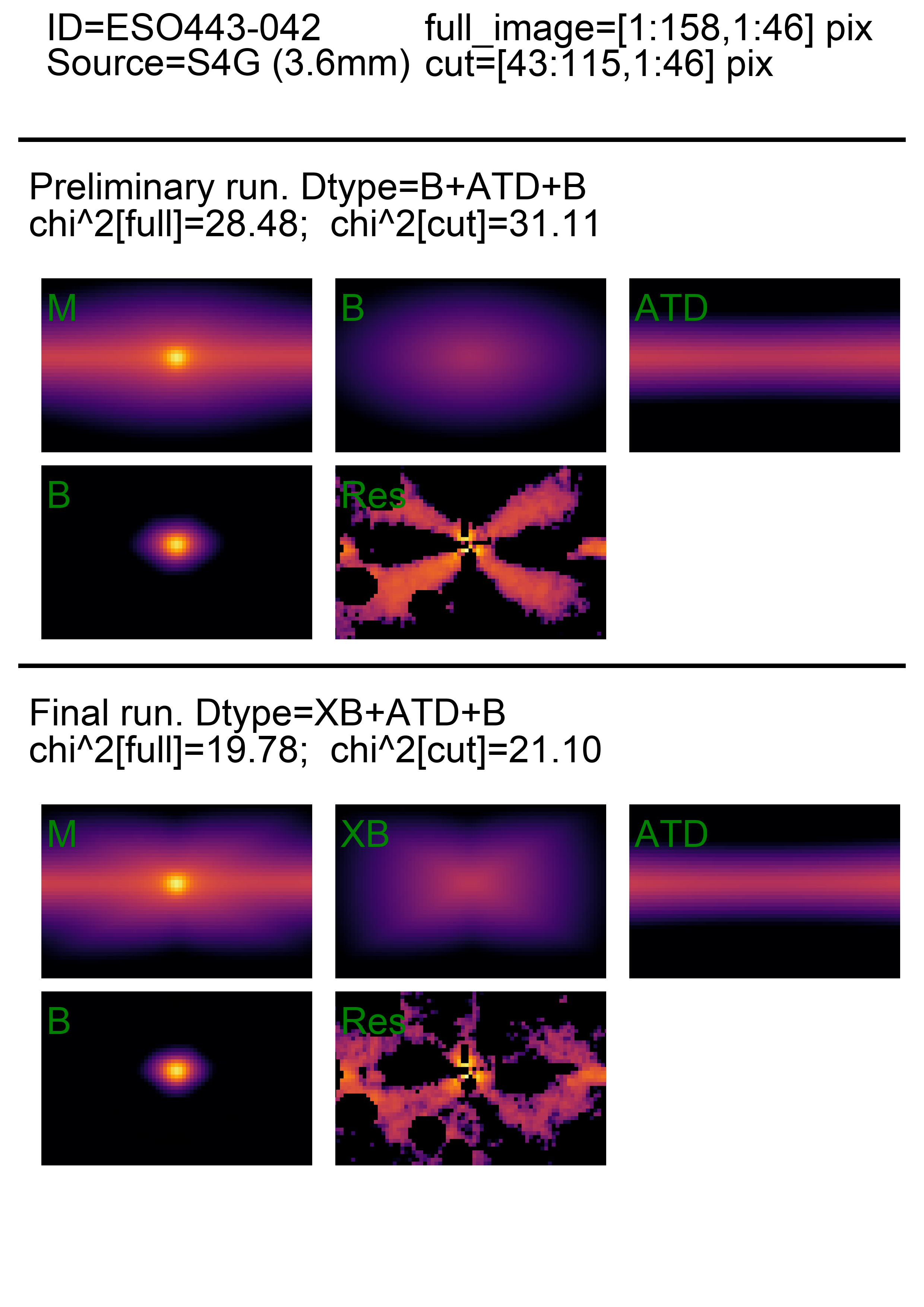}%
\end{minipage}%
\caption{The photometric decomposition of ESO~443-042 galaxy (see caption of Fig.~\ref{fig:NGC128} for details).}
\label{fig:ESO443042}
\end{figure*}
\par
This is a peculiar galaxy where~\cite{Ciambur_Graham2016} identified a peanut with a very small ratio of vertical to in-plane extent, $z_{\Pi,\mathrm{max}}/R_{\Pi,\mathrm{max}}{\approx}0.26$ (in notations of~\cite{Ciambur_Graham2016}) which translates into the angular measure to obtain $\psi=\arctan (z_{\Pi,\mathrm{max}}/R_{\Pi,\mathrm{max}}) {\approx}14^\circ$. At the same time,~\cite{Laurikainen_Salo2017} identified quite a typical B/PS bulge with sizes ratio about $a_\mathrm{X}/b_\mathrm{x}{\approx}0.8$ which corresponds to an angle value $\varphi{\approx}39^\circ$. We note that~\cite{Ciambur_Graham2016} used an $4.5\mu$m image, while~\cite{Laurikainen_Salo2017} used $3.6 \mu$m data. We also use $3.6 \mu$m data. 
\par 
The results of our decomposition for ESO~443-042 is shown in Fig.~\ref{fig:ESO443042}. We masked a bright feature seen
at the left side of the disc (seen as a dark blob at the residue image). It has no symmetrical partner at the other
side of the disc plus it significantly affects the disc component if included. Again, we should deal with two types of
bulges at the preliminary stage. Compared to NGC~128 case, these two bulges are quite distinct in terms of their linear
sizes ($r_\mathrm{X}/r_\mathrm{B}{\approx}23$). Therefore, there is no problem with the possible degeneracy of the
components.
\par
Concerning the X-shaped bulge in this model, we note the following. One can see four rather bright lobes in the
preliminary run residue image. After we transform a Sersic profile into the X-shaped bulge model, these lobes mostly
disperse (although not completely). Chi-square value is substantially reduced too (about one and a half times). However,
it is still rather large ($\chi^2{\approx}21$). Does it mean that our photometric model is no good? From a strict point
of view, yes, it does. However, we should point out that for many galaxies from S4G catalogue encountering such a large
value of chi-square in 2D decomposition is not that unusual (see~\citealt{Salo_etal2015}). In principle, the addition of
other photometric components can help, but in this particular case, we believe that it is unnecessary. The residue image
shows that there are three main sources of errors, namely a central component, right bright blob (although almost
masked), and the residue from the X-shaped bulge. The central component mostly occupies the region with a spatial scale
close to that of the second bulge and therefore should not strongly affect the X-shaped bulge geometric parameters. The
blob clearly lies outside the X-shape bulge area and for this reason also should not significantly affect the X-shaped
bulge model (although it is presence indeed changes the disc model and therefore indirectly affect X-shaped bulge model
but its influence should be dumped by the disc inner truncation).
\par
The geometric parameters of the X-shaped bulge of this galaxy are rather typical. An opening angle $\varphi$ has a value
of about $30^\circ$ and the sharpness parameter value is about $1.3$. The obtained opening angle value is not consistent
with that obtained neither by~\cite{Laurikainen_Salo2016}, nor by~\cite{Ciambur_Graham2016}.  We see no apparent reason
for B/PS bulge identification to produce a very flattened bulge as in~\cite{Ciambur_Graham2016}. Perhaps, strong
inhomogeneities in the disc that we masked could contribute to this matter. As for~\cite{Laurikainen_Salo2017}, their
unsharp-masked image shows two large lobes instead of four separate rays coming from the disc centre (in contrast to our
residue image of the preliminary run shown in Fig.~\ref{fig:ESO443042}). We discuss this matter more thoroughly in
Section~\ref{sec:comp_previous_works}.
\subsubsection{PGC~24926}
\begin{figure*}
\begin{minipage}[t]{0.5 \textwidth}
\includegraphics[scale=0.1]{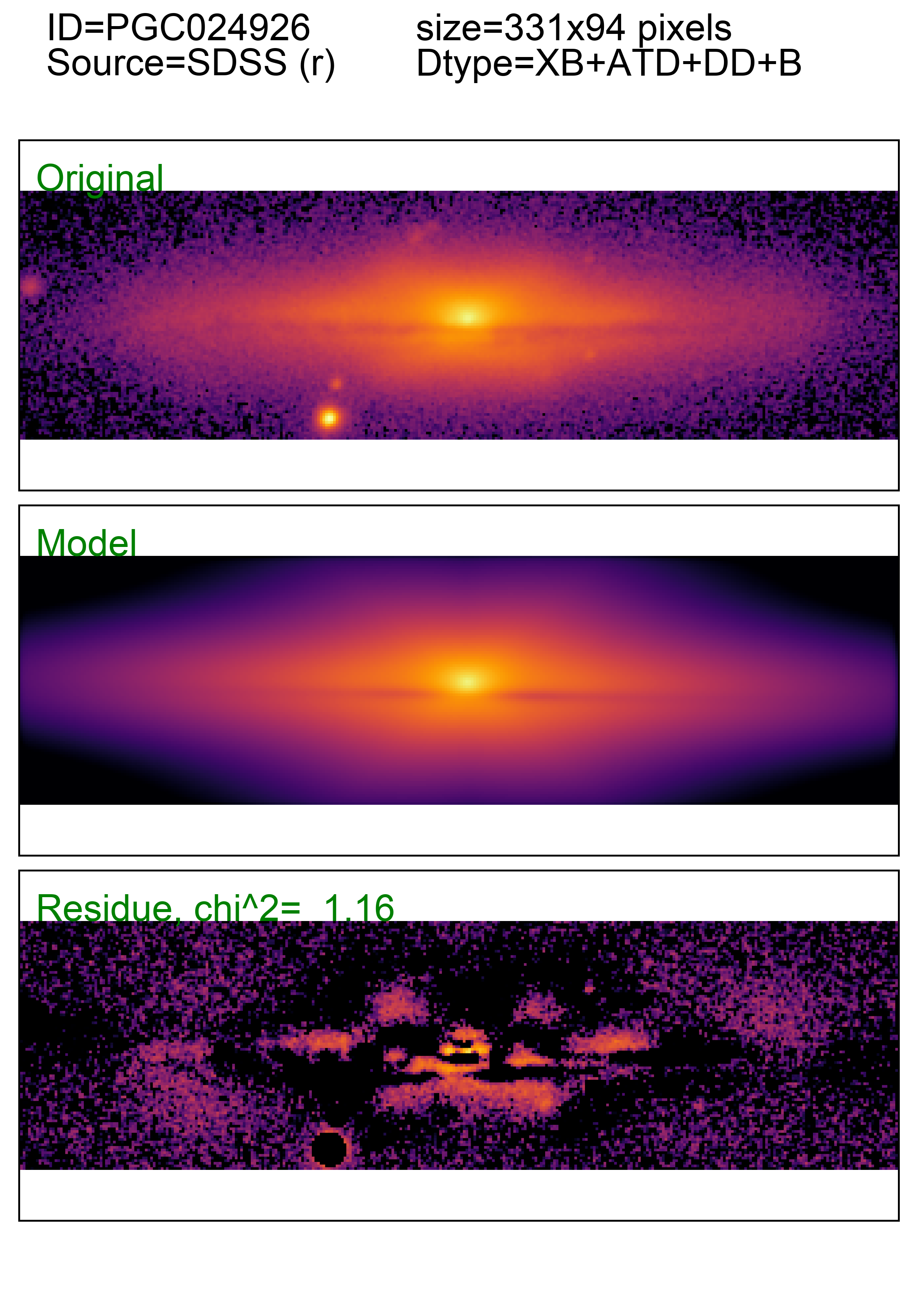}%
\end{minipage}%
\begin{minipage}[t]{0.5 \textwidth}
\includegraphics[scale=0.1]{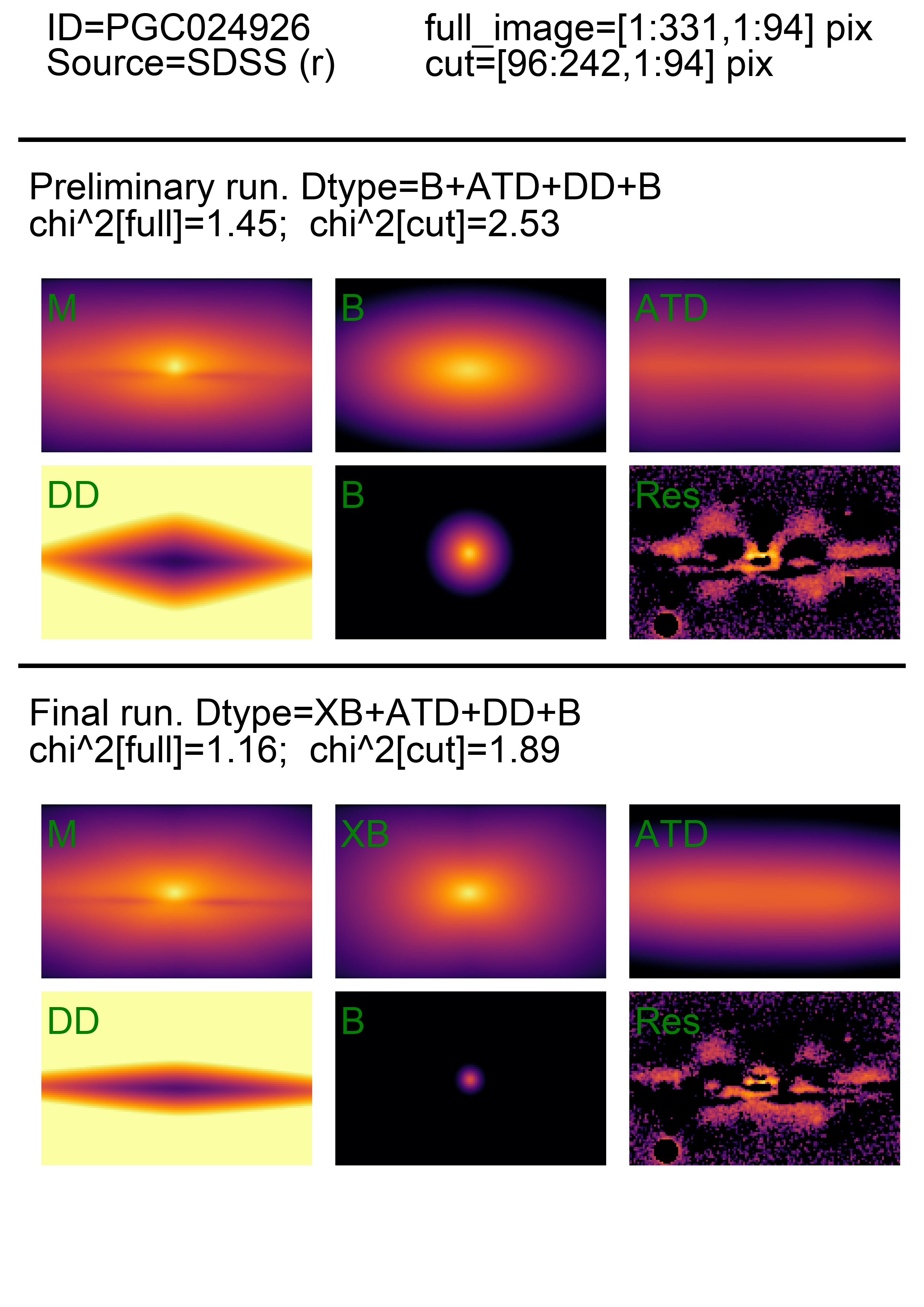}%
\end{minipage}%
\caption{The photometric decomposition of PGC~24926 galaxy (see caption of Fig.~\ref{fig:NGC128} for details)}
\label{fig:103}
\end{figure*}
\par 
The next galaxy which we would like to discuss in more detail is PGC~24926. This galaxy is intersected by an almost
straight dust line \textcolor{black}{visibly shifted from the disc plane}. We show this example to demonstrate how the X-shaped bulge model works in case of a galaxy having a
dust feature. Fig.~\ref{fig:103} shows the result of the decomposition. We can see that the X-shaped profile in residue
images becomes less pronounced from the preliminary to the final run. Although the improvement is not as noticeable as
in previous cases. We note that the second bulge shrinks after modification of the photometric model, while the dust
disc becomes thinner. The resulted peanut has a rather large opening angle $\varphi{\approx}42^\circ$ accompanied by a
small value of the sharpness parameter $r_\mathrm{X}/s_\mathrm{X}{\approx}0.33$. \textcolor{black}{We also note that the centre of the dust disc model turned out to be slightly shifted from the centre of the stellar disc. This effect is probably due to the slight inclination of the stellar disc plane to LoS.}

\subsubsection{Full galaxy sample}
\begin{table}
\centering

\caption{Geometric parameters of X-shaped bulges for full galaxy sample.}

	\centering
\begin{tabular}{c c c c c}
	\hline
	Galaxy&$r_\mathrm{X}$ (kpc)&$\varphi$ (deg)& $r_\mathrm{X}/s_\mathrm{X}$& $\epsilon_\mathrm{X}$\\\hline
	PGC002865&1.38$\pm$0.05&33.86$\pm$0.44&1.75$\pm$0.08&-1.00\\
	PGC010019&2.55$\pm$0.04&29.96$\pm$0.50&1.51$\pm$0.07&-0.71$\pm$0.04\\
	PGC024926&1.15$\pm$0.01&41.75$\pm$0.38&0.33$\pm$0.01&0.40$\pm$0.00\\
	PGC026482&1.31$\pm$0.03&28.45$\pm$0.59&1.15$\pm$0.08&-0.67$\pm$0.06\\
	PGC028788&1.64$\pm$0.04&39.47$\pm$0.82&0.85$\pm$0.09&0.41$\pm$0.03\\
	PGC028900&2.47$\pm$0.06&31.76$\pm$0.37&1.82$\pm$0.06&-1.00\\	
	PGC030221&1.80$\pm$0.08&30.85$\pm$1.19&1.31$\pm$0.38&-0.48$\pm$0.44\\
	PGC032668&1.52$\pm$0.05&29.01$\pm$0.32&1.45$\pm$0.13&-0.61$\pm$0.11\\
	PGC034913&1.71$\pm$0.02&38.14$\pm$1.04&0.46$\pm$0.03&0.06$\pm$0.04\\
	PGC037949&4.53$\pm$0.06&31.34$\pm$0.58&0.62$\pm$0.03&0.37$\pm$0.02\\
	PGC039251&1.97$\pm$0.04&35.02$\pm$0.26&1.47$\pm$0.06&-0.86$\pm$0.02\\
	PGC044422&1.85$\pm$0.03&31.67$\pm$0.32&1.62$\pm$0.04&-0.98$\pm$0.03\\
	PGC045214&2.40$\pm$0.03&32.18$\pm$0.29&1.65$\pm$0.05&-0.89$\pm$0.04\\
	PGC053812&2.96$\pm$0.12&32.33$\pm$0.40&1.30$\pm$0.23&-0.35$\pm$0.27\\
	PGC055959&1.57$\pm$0.09&33.85$\pm$0.53&1.39$\pm$0.11&-0.95$\pm$0.04\\
	PGC069401&2.84$\pm$0.03&42.11$\pm$0.73&0.79$\pm$0.02&0.29$\pm$0.02\\
	PGC069739&2.14$\pm$0.05&39.96$\pm$1.85&0.46$\pm$0.04&0.25$\pm$0.03\\
	ASK361026.0&2.24$\pm$0.02&28.27$\pm$0.22&1.71$\pm$0.03&-1.00\\
	ESO443-042&3.77$\pm$0.21&30.92$\pm$1.48&1.28$\pm$0.15&-0.03$\pm$0.18\\
	NGC128&2.94$\pm$0.03&43.59$\pm$0.36&1.11$\pm$0.02&0.20$\pm$0.01\\
	NGC128$^\star$&2.65$\pm$0.02&44.24$\pm$0.31&0.67$\pm$0.01&0.30$\pm$0.00\\
	NGC2654&1.52$\pm$0.01&34.98$\pm$0.51&0.58$\pm$0.02&0.35$\pm$0.01\\
	NGC2683&1.18$\pm$0.01&31.89$\pm$0.17&0.90$\pm$0.03&-0.40$\pm$0.03\\
	NGC4710&1.23$\pm$0.02&28.73$\pm$0.42&1.12$\pm$0.04&-0.15$\pm$0.05\\
	NGC5529&2.68$\pm$0.08&35.59$\pm$0.91&1.97$\pm$0.15&-0.30$\pm$0.14\\
	NGC7332&0.70$\pm$0.01&38.77$\pm$0.66&0.43$\pm$0.03&0.50$\pm$0.01\\
	\hline
\end{tabular}
\label{tab:xshape_pars}
\end{table}
In Tab.~\ref{tab:xshape_pars} we present the values of X-shaped bulges geometric parameters for our galaxy sample. 
Due to lack of space, we cannot present the decomposition images for all galaxies here. Instead, we made them available at
\url{https://vo.astro.spbu.ru/node/130} and refer an interested reader to look at them online.  \par
 There are several general things we should mention concerning the full galaxy sample. First of all, the X-shaped bulge model works, that is, we indeed could find the reasonable photometric model for almost all galaxies from our sample (with some exceptions mentioned below). But the degree of the residue improvement due to a modification of a simple Sersic model to the X-shaped bulge model varies for different galaxies (see NGC~7332 in online materials for a ``bad'' example). There are also a few cases when we add some other components like a ring or a small central component after the preliminary run. The reason for this is that the ring (or Gaussian) tended to become part of the X-structure if it was included in the preliminary run. It is hard to say whether chi-square improves mainly due to changes in the bulge model or due to the addition of the new component in these cases (see PGC~10019 in online materials). Concerning the disc component, we represented it by an anti-truncated disc model (see Section~\ref{sec:fit}) in all our galaxies. There are several cases when the hole degenerated and the disc showed a plateau instead of an intensity dip. We observed similar behaviour in numerical models in case of a bar having close to end-on orientation. Therefore, it is not surprising that for some real galaxies we observe a similar situation. The matter  of disc inner truncations is interesting by itself, but it is beyond the scope of the present work and we will not discuss it further.
\par
There are several galaxies from~\cite{Savchenko_etal2017} and~\cite{Ciambur_Graham2016} which we discarded after the
preliminary analysis. These are PGC~21357, NGC~678, NGC~2549, NGC~3628, NGC~4111, and NGC~4469. PGC~21357 is a galaxy
with a very strong dust feature. The resulted photometric model which we obtained has an unreasonably small value of the
sharpness parameter ($r_\mathrm{X}/s_\mathrm{X}{\approx}0.3$), with an uncertainty compared with the value itself. This
is due to the strong dust feature screening most of the central part of the X-shaped bulge. NGC~3628 has indeed a very
nice B/PS bulge, but the dust has a very inhomogeneous structure that prevents the reliable disc parameters
estimation. The same story goes for NGC~4469. NGC~678, NGC~2549, NGC~4111 mostly lack an X-shaped feature in
corresponding residue images already at the preliminary stage. Therefore, an X-shaped bulge model tends to become a part
of the disc or just the usual bulge (an X-shaped bulge with $r_\mathrm{X}/s_\mathrm{X}{\approx}0$). Probably, we observe
a bar nearly end-on in these galaxies which explains why they demonstrate boxy-like shape with no apparent X-shape
feature. \par


\section{Comparison with previous works}
\label{sec:comp_previous_works}

\begin{figure}
\center{\includegraphics[scale=0.6]{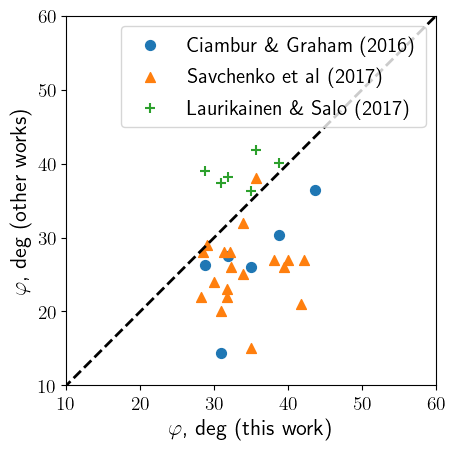}}%
\caption{Comparison of the opening angle values obtained in the previous works and in this work for the same galaxies. For each of the works, with the results of which we try to compare our results, we intersected our galaxy sample with the galaxy samples considered in the mentioned works. The displayed points thus correspond to galaxies from three different sub-samples (which also partially overlap between each other).}
\label{fig:comp_all}
\end{figure}


\begin{figure}
	\center{\includegraphics[width=20cm,
		height=7cm, keepaspectratio]{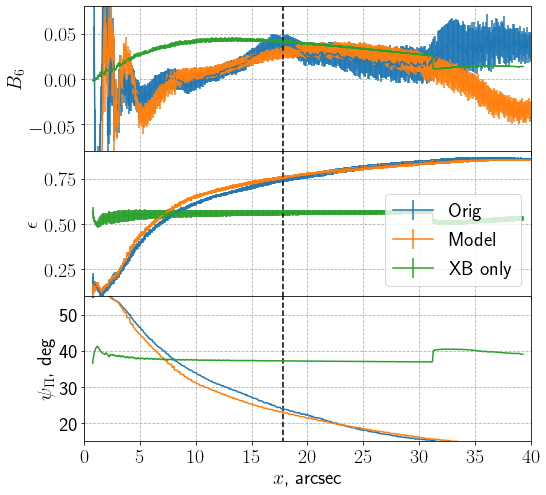}}%
	\caption{Comparison of the isophotes parameters of the original image, the total photometric model for this galaxy obtained in this work, and only the X-shaped bulge component for ESO~443-042 galaxy. \textit{Top}: the dimensionless $B_6$ amplitude profile; \textit{Middle}: the ellipticity of the isophotes $\epsilon$; \textit{Bottom}: the polar angle of the peanut calculated as in~\protect\cite{Ciambur_Graham2016}. Dashed line marks the location of $B_6$ maximum in the original image.}
	\label{fig:comp_CG}
\end{figure}

\begin{figure}

	\center{\includegraphics[scale=0.52]{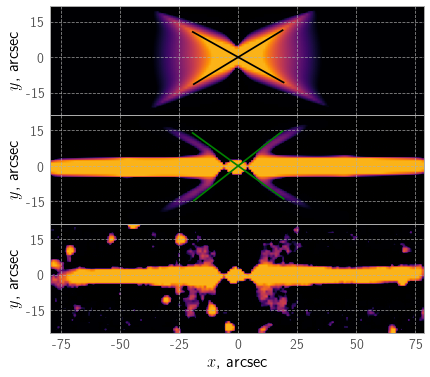}}%

	\caption{Usharp-masked images of the X-shaped bulge component for ESO~443-042 galaxy (\textit{top}), the total
          photometric model (\textit{middle}), and original galaxy image (\textit{bottom}) obtained using the same
          filter parameters and the intensity scales. The black cross highlights the location of the X-structure rays. \textcolor{black}{The green cross (\textit{middle panel}) highlights the location of the X-structure rays if they had an opening angle about $37^\circ$, which is the value measured by ~\protect\cite{Laurikainen_Salo2017} for this galaxy.}}
	\label{fig:comp_LS}
\end{figure}

\begin{figure}
\includegraphics[scale=0.6]{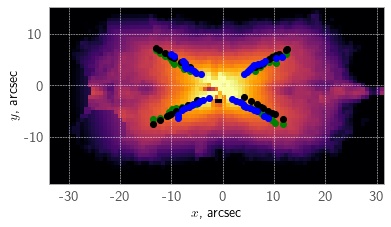}%
\caption{Comparison of approaches to measure X-structure opening angles in the present work and in~\protect\cite{Savchenko_etal2017} for ESO~443-042. Green and blue dots correspond to the location of the density peaks identified along the elliptical cuts with ellipticities $\epsilon=0.4$ and $\epsilon=0.6$, respectively. Black dots mark the actual location of the X-structure rays in the X-shaped bulge model for this galaxy.}
\label{fig:comp_S}
\end{figure}

Fig.~\ref{fig:comp_all} shows how the opening values obtained in this work correlate with the values obtained
by~\cite{Ciambur_Graham2016},~\cite{Laurikainen_Salo2017} and~\cite{Savchenko_etal2017} for the same galaxies. The
overall picture seems rather unsatisfactory. In an ideal case, if the values do not depend on the method used, we
should obtain strictly equal values. However, this is not true for most of the considered galaxies. Rather, they try
to avoid the line of equal values. There are also some trends that we observe in the figure. The opening angles
from~\cite{Laurikainen_Salo2017} are larger on average than those we measure. The opposite situation is
with~\cite{Ciambur_Graham2016} and~\cite{Savchenko_etal2017}. Their values are smaller than ours on average.
\par
 Although it is not that rare that different methods produce different results for the same galaxies, we would like to get a general idea of why this happens and happens in the manner shown in Fig.~\ref{fig:comp_all}. To answer this question, we analyse in more detail how different measuring techniques work on the example of already discussed ESO~443-042 galaxy. 
\par
To compare our results with the results of~\cite{Ciambur_Graham2016} we apply the isophotal analysis similar to that used by~\cite{Ciambur_Graham2016} to the original galaxy image and the images of the total photometric model obtained in this work, and only the X-shaped bulge component extracted by 2D photometric decomposition. In Fig.~\ref{fig:comp_CG} we plot the main parameters of the isophotes that~\cite{Ciambur_Graham2016} used to obtain the peanut angle. These are $B_6$ harmonic amplitude and the ellipticity of the isophotes $\epsilon$. The peanut angle is calculated as follows:
\begin{equation}
\psi_\Pi = -\arctan \left[  (\epsilon_\mathrm{max}-1)\tan (2 \pi/6)\right],
\label{eq:peanut_angle}
\end{equation}
where $\epsilon_\mathrm{max}$ is the ellipticity of the isophote that has a maximum $B_6$ value\footnote{We note that the equation we give here is different from that used by~\cite{Ciambur_Graham2016} for $\psi_\Pi$ evaluation. We assume that there is probably a typo in their equation, because $\psi_\Pi$ they indicated in their figure~1 is not the eccentric anomaly angle, but the polar angle. The eccentric anomaly for the peanut should be $2\pi/6$, while the polar angle should indeed depend on the isophotes ellipticity.}. As can be seen from Fig.~\ref{fig:comp_CG}, $B_6$ profiles and the ellipticity profiles are close for the original image and for the total photometric model. However, when we apply isophotal analysis only to the X-shaped bulge component, we obtain drastically different results. Here observe a constant value of the isophotes ellipticity, and, moreover, it is considerably greater than that of the original image or the best-fit photometric model in the inner area and far smaller than those in the outer area. At $B_6$ maximum location $\epsilon$ has the value of about 0.56 for the X-shaped bulge component, while in the original image it is about 0.77. According to that, the peanut angle values are also quite different. In the former case, we obtain the value of about 38$^\circ$, while in the second case the angle value is about 21$^\circ$. Clearly, the only culprit responsible for the observed difference is a disc component. In fact, Fig.~\ref{fig:comp_CG} directly indicates that the peanut angle value obtained based on $B_6$ harmonic is highly dependent on the disc contribution. With the addition of a disc to the B/PS bulge, the isophotes become more elongated, the ellipticity decreases and, according to Eq.~(\ref{eq:peanut_angle}), the angle value decreases too. We assume that the observed systematic shift of~\cite{Ciambur_Graham2016}'s angles to lower values for other galaxies can also be explained in a similar way.  
\par
Concerning the comparison with~\cite{Laurikainen_Salo2017}'s results, we should state that the strict comparison is
hardly possible as the usharp-masking procedure has some parameters that determine the outcome of the procedure. These
parameters usually depend on the particular galaxy, are chosen by the observer and not specified.  To understand more
clearly how the unsharp-masking technique works for the B/PS bulge and their X-structures, we prepared the
unsharp-masked images of the X-shaped bulge model, the total photometric model, and the original ESO~443-042 image using
the same filter parameters and the intensity scales in both cases (Fig.~\ref{fig:comp_LS}). As can be seen from the
figure, for the X-shaped model we find that unsharp-masking indeed highlights the locations of the X-structure rays (and
quite accurately on top of it). However, in the case of a full galaxy image, the unsharp-masking transforms the whole bulge
into two ear-like structures with the large intensity dip between them. The X-structures are less noticeable on the
original image (the two left rays in particular), although they still can be identified. The next question is how we
measure the parameters of these lobes. As far as we understand, \cite{Laurikainen_Salo2017} measured them by hand, and
this can introduce some errors, which result in the systematic shift of the opening angles. We also note that while
experimenting with unsharp masking, we found that increasing the gradient contribution as much as possible, but to such
a degree that the density peaks are still visible, helps to more accurately identify the X structure.
\par 
For completeness, we compare our approach with the approach used by~\cite{Savchenko_etal2017} on the example of
ESO~443-042 galaxy although~\cite{Savchenko_etal2017} did not consider this galaxy in their sample. This can be done if
we analyse the density distribution along the photometric cuts as it was done in~\cite{Savchenko_etal2017}. One of the
interesting aspects of that study was that the authors considered cuts not of the B/PS bulge model itself but of the
image obtained by subtracting all components from the original image, except the one responsible for the B/PS bulge.  We
prepared such type of an image for ESO~443-042 (Fig.~\ref{fig:comp_S}) and analysed the density distributions along the
ellipses with different major axes values in a range from about 8 to 18 arcsec. We consider two sets of ellipses one
with the ellipses ellipticity $\epsilon=0.4$ (green dots) and the other one with $\epsilon=0.6$ (blue dots). Along each
of the ellipses, we identified four density peaks that correspond to the points where the rays of the X-structure
intersect with the ellipses. For reference, we also added the set of points that highlight the location of the
ray in our X-shaped bulge model for this galaxy (black dots). As can be seen, the location of all density peaks coincide
remarkably well for all rays except the bottom right one but the difference there is not that large. The consistency
that we observe directly indicates that the photometric cuts approach should produce results close to ours if other
things are the same. According to that, there should be another reason for the systematic shift that we observe
in~Fig.~\ref{fig:comp_all}. In Section~\ref{sec:numerical_models}, where we studied the impact of different disc models
on the X-shaped bulge model, we showed that the application of a simple exponential disc model (without an inner
truncation) lead to lower values of the opening angles with the effect stronger for close to end-on bars (see
Fig.~\ref{fig:compare_discs}).~\cite{Savchenko_etal2017} indeed used an exponential disc model while we use AT disc
model. Consequently, it is logical to associate the observed systematic shift with different disc models applied.
\section{Comparison with the modelled galaxies}
\label{sec:comp_models}

\begin{figure}
\begin{minipage}[t]{0.5 \textwidth}
\includegraphics[scale=0.5]{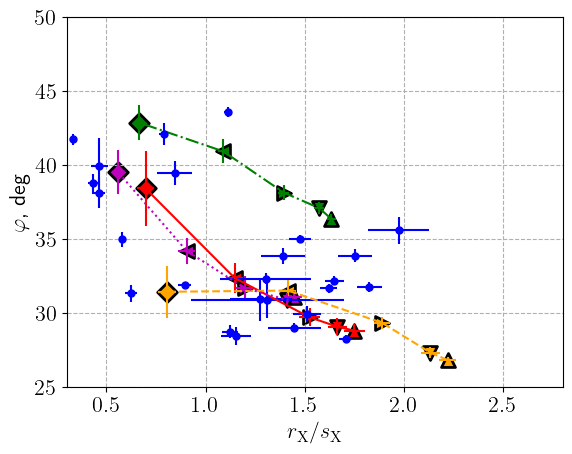}%
\end{minipage}%
\caption{Comparison of opening angles and the sharpness parameters of the X-shaped bulges of real (blue dots) and simulated (triangles and hexagons) galaxies from Fig.~\ref{fig:compare_models}.}
\label{fig:real_galaxies_phi_sharpness}
\end{figure}

Here we compare the modelled and real galaxies in terms of their X-shaped bulge parameters. Fig.~\ref{fig:real_galaxies_phi_sharpness} shows the data obtained for real galaxies superimposed on the data obtained for the modelled galaxies from Fig.~\ref{fig:compare_models}. We can see that the real galaxies mostly follow the modelled ones in terms of their B/PS bulges. \textcolor{black}{In terms of opening angles real X-structures span the range from about $27^\circ$ to $44^\circ$. We note that the lower boundary value $27^\circ$ translates into the aspect ratio about $0.51$ which agrees quite well with the lower boundary found by~\cite{Laurikainen_Salo2017} for both real galaxies and $N$-body models they used (see their figure 8).} We can also note that for the most of real B/PS bulges the sharpness parameter lies in the range from $r_\mathrm{X}/s_\mathrm{X}{{\approx}}1$ to $r_\mathrm{X}/s_\mathrm{X}{\approx}2$. The modelled bulges demonstrate  similar values of the sharpness parameter for bars \textcolor{black}{rotated from about $45^\circ$ to $90^\circ$ around the disc axis}. There are also some galaxies the bulges of which have the parameters that do not fall on the model curves. For example, the bulge in NGC~2654 galaxy has an opening angle value  of its X-structure $\varphi{\approx}30^\circ$ and a rather small value of the sharpness parameter, $r_\mathrm{X}/s_\mathrm{X}{\approx}0.6$ (the point which is the closest to the bottom left corner of the graph). 
There is also NGC~5529 galaxy where the X-shaped bugle has $\varphi{\approx}36^\circ$ and the sharpness parameter value $r_\mathrm{X}/s_\mathrm{X}{\approx}2$. The corresponding point is the rightmost one. Contrary to the usual train of thoughts, this is rather reassuring that there are some bulges that are not consistent with the modelled bulges in terms of their parameters. This is precisely because we know that our models should not describe all possible cases and especially considering the fact the model curves do not account for the evolution of the B/PS bulges in time. In fact, Fig.~\ref{fig:comp_all} encourages us to consider more of different galaxy models at different time moments which we plan to do in future.
\par 
 Another important observation that we can make from Fig.~\ref{fig:real_galaxies_phi_sharpness} is the following. The figure shows that we do not detect real B/PS bulges with X-structure angles smaller than those we obtained in the modelled galaxies. Thus, the problem with small opening angles in real galaxies which we encountered in our previous work~\citep{Smirnov_Sotnikova2018} can be considered solved. We can conclude the discrepancy between model and real data which was observed in that work is due to the difference in measuring approaches and there is no need to find some physical reasons for it. 
\par


\section{Summary and conclusions}
\label{sec:sum}
Concluding our work, we can state the following:
\begin{enumerate}
	\item  We introduced a new photometric model accounting simultaneously for the boxy/peanut-shaped (B/PS) bulges and the X-structures that reside in such bulges. The functional form of the model is that of a 2D Sersic function with a truncated intensity profile. The truncations are introduced above (in the upper half-plane) and below (in the bottom half-plane) the rays of X-structures. The model allows one to measure the geometric properties of X-structures (opening angle, linear scale) by means of photometric decomposition. We call this new model ``the X-shaped bulge model''.
	%
	\item The form of the model was deduced based on the appearance of the B/PS bulge in some particular numerical model of the galaxy where such a bulge naturally arises due to the bar thickening. The considered type of numerical model is frequently used to study the secular evolution of early-type SB galaxies. An important aspect of our study is that we first extracted the B/PS bulge using some dynamical considerations. In particular, we split the modelled galaxy into different dynamical subsystems (bar + disc) using the frequency characteristics of particles-``stars''. This allowed us to study the B/PS bulge density distribution in its ``pure'' form without the need to take into account the disc component when we try to find an appropriate photometric model for the B/PS bulge.
	\item Using a set of different galaxy models, we study how the parameters of the X-shaped bulge photometric model vary depending on the projection effects (disc inclination, bar \textcolor{black}{azimuthal viewing angle}), the applied photometric model of the disc (a simple exponential disc and an anti-truncated disc) and in case of co-existence of two bulges in the same galaxy. We found that the application of the anti-truncated disc model leads to more reasonable values of the X-shaped bulge parameters than the application of a simple exponential disc model for all disc inclinations and bar orientations considered. In case of the co-existence of two bulges (a B/PS bulge and a classic one), we found that it is indeed possible to retrieve ``true'' values of the X-shaped bulge parameters even if the bulges are superimposed onto each other.  
	%
	\item We applied the X-shaped bulge model to study the B/PS bulges of some real galaxies. In total, we considered 29 galaxies. The B/PS bulges of these galaxies were previously studied by~\cite{Ciambur_Graham2016},~\cite{Savchenko_etal2017} and~\cite{Laurikainen_Salo2017} (in each work some sub-sample of our sample) using different methods. For three of these galaxies (NGC~678, NGC~2549, and NGC~4111), we could not obtain a reliable photometric model with the X-shaped bulge component. This is due to the fact that these galaxies did not show a bright X-shape feature when we subtracted Sersic bulge and disc components.  We also could not obtain the reliable photometric model for PGC~21357, NGC~3628, and NGC~4469 due to dust feature in these galaxies. For each of the remaining 25 galaxies, we obtained the best-fit photometric model, which includes an X-shaped bulge model as a separate component. We compared the obtained values of opening angles with results of~\cite{Ciambur_Graham2016},~\cite{Savchenko_etal2017} and~\cite{Laurikainen_Salo2017} for the same galaxies. We found that the application of different measuring techniques leads to systematically different results on exactly the same galaxies. On the example of ESO~443-042 galaxy, we showed that 6-th Fourier harmonic amplitude $B_6$ used by~\cite{Ciambur_Graham2016} to characterise B/PS bulges is highly dependent on the disc contribution, and the peanut angle obtained based on it does not correspond to the X-structure opening angle in some cases.
	\item We compared the geometric characteristics of B/PS bulges of some simulated galaxies with those obtained for real galaxies. We found that the values are generally consistent with each other if we measure them in exactly the same way for simulated and real galaxies. We found that the real X-structures do not show opening angles smaller than about 27$^\circ$. Thus, we resolved an inconsistency between the modelled X-structures and the X-structures of real galaxies we encountered in our previous work~\citep{Smirnov_Sotnikova2018}. In that work, the X-structures of modelled galaxies showed opening angles on average larger than those of real galaxies. In the present work, we explicitly showed that the difference is due to different approaches to the X-structure measuring. 
\end{enumerate}
Concerning future plans, there are two different lines of work we can pursue. First, we did not consider how B/PS bulge parameters evolve with time in the present work, and it is quite an important thing to check since the bars can buckle, and it is possible that using our new model we can capture such a behaviour. We also did not consider how the B/PS bulges of different types of bars, for example, a bar with a barlens or a peanut-shaped bar if viewed face-on, differ in terms of their parameters, which is an interesting thing to check. \textcolor{black}{Two mentioned types were already compared in~\cite{Laurikainen_Salo2017} in terms of the B/PS bulges linear sizes but no significant difference was found. Perharps, using our new model we can reveal some additional differences between these two cases.} The second important line is the expansion of the sample of real galaxies and a more thorough comparison of the parameters of real and modelled galaxies. 
\par
As a concluding remark, we would like to note the following. The photometric model deduced here is not absolute, in a sense that a more accurate model can be obtained in the future. For example, there can be a model that accounts for the thickness of X-structure rays in an explicit way or the model that somehow traces the bending of its rays. Obviously, the end goal of this road is to find an appropriate 3D bar model, which includes the X-structures as its natural part. The present work can be viewed as the stepping stone on this road.

\section*{Data availability}
The data underlying this article will be shared on reasonable request to the corresponding author.

\section*{Acknowledgements}
The authors express gratitude for the grant of the Russian Foundation for Basic Researches number 19-02-00249. AS also thanks N.~Ya.~Sotnikova and A.~A.~Marchuk for a helpful discussion of some aspects of this work.

\bibliographystyle{mnras}
\bibliography{main}

\bsp	
\label{lastpage}
\end{document}